\documentclass
[showkeys,showpacs,superscriptaddress,preprintnumbers,byrevtex]{revtex4}
\usepackage{graphicx}
\usepackage{bm}
\begin{document}
\preprint{CYCU-HEP-11-03}
\title{Chiral restoration at finite $T$ under the magnetic field with the meson-loop corrections}
\author{Seung-il Nam}
\email[E-mail: ]{sinam@kau.ac.kr}
\affiliation{Research Institute for Basic Sciences, Korea Aerospace University, Goyang 412-791, Korea}
\author{Chung-Wen Kao}
\email[E-mail: ]{cwkao@cycu.edu.tw}
\affiliation{Department of Physics, Chung-Yuan Christian University, Chung-Li 32023, Taiwan}
\date{\today}
\begin{abstract}
We investigate the (partial) chiral restoration at finite temperature $(T)$ under the strong external magnetic field $\bm{B}=B_{0}\hat{z}$ of the SU(2) light-flavor QCD matter.
To this end, we employ the instanton-liquid QCD vacuum configuration accompanied with the linear Schwinger method for inducing the magnetic field. The Harrington-Shepard caloron solution is used to modify the instanton parameters, i.e. the average instanton size $(\bar{\rho})$ and inter-instanton distance $(\bar{R})$, as functions of $T$. In addition, we include the meson-loop corrections (MLC) as the large-$N_{c}$ corrections because they are critical for reproducing the universal chiral restoration pattern. We present the numerical results for the constituent-quark mass as well as chiral condensate which signal the spontaneous breakdown of chiral-symmetry (SB$\chi$S), as functions of $T$ and $B_{0}$. From our results we observe that the strengths of those chiral order parameters are enhanced with respect to $B_{0}$ due to the magnetic catalysis effect. We also find that there appears a region where the $u$ and $d$-quark constituent masses coincide with each other at $eB_0\approx(7\sim9)\,m^2_\pi$, even in the presence of the explicit isospin breaking ($m_u\neq m_d$). The critical $T$ for the chiral restoration $T_c$ tends to shift to the higher temperature in the presence of the $B_0$ for the chiral limit but keeps almost stationary for the physical quark mass case. The strength of the isospin breaking between the quark condensates is also explored in detail by defining the ratio $\mathcal{R}\equiv(\langle iu^\dagger u\rangle-\langle id^\dagger d\rangle)/(\langle iu^\dagger u\rangle+\langle id^\dagger d\rangle)$, which indicates the competition between the explicitly isospin breaking effect and magnetic catalysis effect. We also compute the pion weak-decay constant $F_\pi$ and pion mass $m_\pi$ below $T_c$, varying the strength of the magnetic field, showing correct partial chiral restoration behaviors. Besides we find that the changes for the $F_\pi$ and $m_\pi$ due to the magnetic field is relatively small, in comparison to those caused by the finite $T$ effect.
\end{abstract}
\maketitle
\section{Introduction}
Studies on the breakdown of symmetries and their restorations have been very useful in the analysis of phenomena related to phase transitions. Such studies applied to {\it Quantum Chromodynamics} (QCD) are particularly fascinating since QCD owns a very complicated phase structure. Among them, the (partial) restoration of chiral symmetry at finite temperature ($T$) and/or quark chemical density ($\mu$) has been one of the most interesting and stimulating subjects for decades. A great number of theoretical and experimental endeavors have been devoted to this subject.
In particular, the high-$T$ and low-$\mu$ region, which resembles the early universe, shows a very noble feature of the restoration pattern.
Namely there is a crossover phase transition for the nonzero current-quark mass ($m_{f}$) and that of the second-order for $m_{f}=0$. These distinctive patterns of the phase transition are consistent with the universal-class argument of the three-dimensional Ising model~\cite{de Forcrand:2003hx}, and turns out to be highly nontrivial in QCD~\cite{Aoki:2006we,Endrodi:2011gv}. Moreover, the critical endpoint (CEP)~\cite{Gavai:2004sd,Fujii:2003bz}, tri-critical point (TCP)~\cite{Schaefer:2006ds}, the critical chiral phase transition on the $T$-$\mu$ plane have been also attracting much interest.

Recently, together with the energetic progress for the heavy-ion-collision (HIC) experiment facilities, such as the relativistic heavy-ion collider (RHIC) and large hadron collider (LHC), one can now probe hot and dense QCD matters, i.e. {\it quark-gluon plasma} (QGP), experimentally.
It has been reported that very strong magnetic field in the order of the several times of $m^{2}_{\pi}\,[\mathrm{GeV}^{2}]$ can be produced in the noncentral (peripheral) heavy-ion collision (HIC) experiment by STAR collaboration at RHIC~\cite{:2009txa,:2009uh}. According to this strong magnetic field and $CP$-violating domains created inside the QGP, signals for possible $P$ and $CP$ violations were observed as the charge separation along the direction of the external magnetic field, which is perpendicular to the collision plane. Theoretically, this phenomenon is nothing but the axial anomaly of electromagnetic currents~\cite{Kharzeev:2004ey,Voloshin:2004vk,Kharzeev:2007jp,Voloshin:2008jx,:2009txa,:2009uh,Fukushima:2008xe,Fukushima:2009ft,Warringa:2009rw,Asakawa:2010bu}. The charge separation at relatively low $T$ has been already investigated within the instanton-vacuum framework by one of the authors (S.i.N.)~\cite{Nam:2009jb,Nam:2009hq,Nam:2010nk}, in which related works and references can be found. Actually even before this sort of studies receiving much more interest recently due to the energetic progress of the HIC physics, QCD under magnetic field had been an important subject ~\cite{Menezes:2008qt,Boomsma:2009yk,Gatto:2010qs,Mizher:2010zb,Nam:2010mh}. Inside QCD matter in the presence of the external magnetic field, the spins of the quarks are aligned along the direction of induced magnetic field according to their helicities. As a result, the quark-antiquark pair couples strongly, i.e. which is a phenomena denoted by the {\it magnetic catalysis}~\cite{Boomsma:2009yk,Miransky:2002rp}. Hence, taking into account that the order parameter of the spontaneous breakdown of chiral symmetry (SB$\chi$S) is the chiral (quark) condensate $\langle \bar{q}q\rangle $, one can  observe that SB$\chi$S will be enhanced by the presence of the external magnetic field. Accordingly there appear specific consequences as 1) the enhancement of the critical $T$ for SB$\chi$S, $T_c$, 2) the increase of constituent-quark mass, and 3) modifications in the low-energy constants (LEC). In the present work, we will address to all of these interesting consequences. To date, there have been many related works, for instance, from the Nambu-Jona-Lasinio (NJL) model~\cite{Menezes:2008qt,Boomsma:2009yk}, holographic QCD~\cite{Filev:2009xp,Callebaut:2011uc}, lattice QCD simulations~\cite{D'Elia:2011zu}, linear sigma model~\cite{Mizher:2011wd}, Polyakov-loop inspired model~\cite{Ruggieri:2011qy}, and so on.

In the present work, we want to investigate the (partial) chiral restoration under the strong external magnetic field in QCD matter. For this purpose, we employ the  instanton-liquid model, modified by the Harrington-Shepard caloron solution finite $T$~\cite{Harrington:1976dj}. Although this approach does not manifest the quark confinement, such as the nontrivial holonomy caloron, i.e. the Kraan-van Baal-Lee-Lu caloron~\cite{Kraan:1998pm,Lee:1998bb},  as shown in Refs.~\cite{Nam:2009nn,Nam:2009jb,Nam:2009hq,Nam:2010nk,Nam:2010mh}, it is a useful nonperturbatve method to study QCD matter at finite $T$. To include the external magnetic field, we make use of the linear Schwinger method~\cite{Schwinger:1951nm,Nam:2009jb,Nam:2009hq}. Besides we also take into account the meson-loop corrections (MLC) for the SU(2) light-flavor sector as the large-$N_c$ corrections. MLC is essential to reproduce the correct current-quark mass dependence of relevant physical quantities~\cite{Nam:2008bq} and universal-class chiral restoration pattern at finite $T$~\cite{Nam:2010mh}.

Our numerical results show that the chiral order parameters, such as the constituent-quark mass and chiral condensate, are enhanced with respect to $B_0$ due to the magnetic catalysis effect. There is a region where the $u$- and $d$-quark constituent masses coincide at $eB_0\approx(7\sim9)\,m^2_\pi$, even for the explicit isospin-symmetry breaking, i.e. $m_u\ne m_d$. The critical $T$ for the chiral restoration, $T_c$ tends to be shifted higher pronouncedly in the presence of the $B_0$ in the chiral limit. On the contrary $T_{c}$ keeps
almost stationary for the physical quark mass case.
The strength of the isospin breaking between the $u$ and $d$ quark condensates is also explored in detail by defining the ratio $\mathcal{R}\equiv(\langle iu^\dagger u\rangle-\langle id^\dagger d\rangle)/(\langle iu^\dagger u\rangle+\langle id^\dagger d\rangle)$ as a function of $T$ and $B_0$. Finally, we compute the pion weak-decay constant $F_\pi$ and pion mass $m_\pi$ below $T_c$ as functions of $T$ and $B_0$, showing correct partial chiral restoration behaviors. Our result also shows the changes of the $F_\pi$ and $m_\pi$ due to the magnetic field, are relatively small in comparison to those caused by the finite $T$ effect.

The present work is structured as follows: In Section II, we make a brief introduction of the basic instanton-liquid model at vacuum without MLC and explain typical procedures to compute relevant quantities for further discussions. In Section III,  we consider the inclusion of the MLC and the external magnetic field. The $T$-dependent modification of the instanton parameters using the Harrington-Shepard caloron is given in Section IV. Taking into account all the ingredients in the previous Sections, we derive the expressions for the saddle-point equation and chiral condensate as functions of $T$ ad $B_0$ in Section V. Section VI is devoted to presenting numerical results and associated discussions including the analysis of the pion properties at finite $T$ and $B_0$. Summary and conclusion are given in Section VII.
\section{E$\chi$A in the leading order of the large $N_{c}$ expansion in vacuum}
We start by making a brief introduction for the instanton-liquid model in vacuum. This theoretical framework is characterized by the average of the inter-(anti)instanton distance $\bar{R}\approx1$ and that of the (anti)instanton size $\bar{\rho}\approx1/3$ fm~\cite{Diakonov:1985eg,Diakonov:1995qy}. The effective chiral action (E$\chi$A) in the leading order (LO) of the $1/N_{c}$ expansion can be written in Euclidean space as follows:
\begin{eqnarray}
\label{eq:EA1}
\mathcal{S}_{\mathrm{eff}}[m_{f}]&=&\mathcal{C}
+\mathcal{N}\ln\lambda+2\sigma^{2}
-\int\frac{d^4k}{(2\pi)^4}\mathrm{Tr}_{c,f,\gamma}
\ln\left[\frac{\rlap{/}{k}+i[m_{f}+M(k)] }{\rlap{/}{k}+im_{f}}\right],
\end{eqnarray}
where $\mathcal{C}$, $\mathcal{N}$, $\lambda$, $\sigma$, and $m_{f}$ correspond to an irrelevant constant for further discussions, the instanton number density (packing fraction), the Lagrangian multiplier to exponentiate the $2N_{f}$-'t Hooft interaction, the saddle-point value of the chiral condensate, and current-quark mass for the flavor $f$ for the SU(2) light-flavor sector. The $\mathrm{Tr}_{c,f,\gamma}$ indicates the trace over the color, flavor, and Lorentz indices. Detailed explanations on these instanton-related quantities are given in Refs.~\cite{Diakonov:1985eg,Diakonov:1995qy}. In this picture quarks are moving inside the (anti)instanton ensemble and flipping their helicities. It results in that (anti)quarks acquire the momentum-dependent effective masses dynamically, i.e. constituent-quark masses. Assuming that the zero modes dominate the low-energy phenomena, we can write the Dirac equation for a quark for the (anti)instanton background as follows:
\begin{equation}
\label{eq:ZERO}
\left(i\rlap{/}{\partial}+\rlap{\,/}{A}_{I\bar{I}} \right)\Phi_{I\bar{I}}=0,
\end{equation}
where $A_{I\bar{I}}$ and $\Phi_{I\bar{I}}$ denote respectively the singular-gauge (anti)instanton solution and eigen function of the Dirac equation in the coordinate space~\cite{Diakonov:1985eg}. By performing Fourier transformation of the $\Phi_{I\bar{I}}$, one is led to a momentum-dependent effective quark mass:
\begin{equation}
\label{eq:MDEQM}
 M(k)\equiv M_{a}=M_{f}F^{2}(k),\,\,\,\,
F(k)=2t\left[I_{0}(t)K_{1}(t)-I_{1}(t)K_{0}(t)-\frac{1}{t}I_{1}(t)K_{1}(t) \right],\,\,\,\,t\equiv\frac{|k|\bar{\rho}}{2},
\end{equation}
where $M_{f}$ stands for the constituent-quark mass for each flavor $f$,  $K_{n}$ and $I_{n}$ are the modified Bessel functions~\cite{Diakonov:1995qy}. Note that the $F(k)$ can be interpreted as a quark distribution and plays the role of a natural UV regulator. Hence, in the instanton approach, UV divergences are regularized by construction without inserting any artificial form factors by hand. In practice, it is much easier to employ a parametrized form of the $F(k)$ as in Refs.~\cite{Nam:2009jb,Nam:2009hq,Nam:2010nk}:
\begin{equation}
\label{eq:FFPARA}
F(k)=\frac{2}{2+k^{2}\bar{\rho}^{2}}.
\end{equation}

From the E$\chi$A we derive the following self-consistent (saddle-point) equations. They are used to determine relevant quantities such as the constituent-quark mass $M_{f}$ at zero-momentum transfer $k^2=0$ in Eq.~(\ref{eq:MDEQM}):
\begin{equation}
\label{eq:SCE}
\frac{\partial\mathcal{S}_{\mathrm{eff}}[m_{f}]}{\partial\lambda}=0,
\,\,\,\,
\frac{\partial\mathcal{S}_{\mathrm{eff}}[m_{f}]}{\partial\sigma}=0,
\end{equation}
Similarly the chiral condensate can be computed by differentiating the E$\chi$A with respect to $m_{f}$:
\begin{equation}
\label{eq:CC}
-\frac{1}{N_{f}}
\frac{\partial\mathcal{S}_{\mathrm{eff}}[m_{f}]}{\partial m_{f}}
=\langle iq^{\dagger}q\rangle .
\end{equation}
From the first equation in Eq.~(\ref{eq:SCE}), and Eq.~(\ref{eq:CC}), one obtains the expressions for the LO contributions for the instanton number density and the chiral condensate as functions of $m_{f}$ as follows:
\begin{equation}
\label{eq:LO0}
\mathcal{N}_{\mathrm{LO}}
=2N_{c}N_{f}\int\frac{d^4k}{(2\pi)^4}
\left[\frac{M_{a}\bar{M}_{a}}
{k^{2}+\bar{M}^{2}_{a}}\right],\,\,\,\,
\langle iq^{\dagger}q\rangle_{\mathrm{LO}}
=4N_{c}\int\frac{d^4k}{(2\pi)^4}
\left[\frac{\bar{M}_{a}}{k^{2}+\bar{M}^{2}_{a}}
-\frac{m_{f}}{k^{2}+m^{2}_{f}} \right].
\end{equation}
Here we define $\bar{M}_{a}=m_{f}+M_{a}$. The value of ${\cal N}_{LO}$ is determined from the parameter $\bar{R}$ and its phenomenological value is $ \sim(200\,\mathrm{MeV})^{4}$~\cite{Diakonov:1995qy}.
At the chiral limit $m_f=0$, one solves the first equation of Eq.~(\ref{eq:LO0}) self-consistently. The value of $M_{f}$ turns out to be about $325$ MeV. It is well consistent with the constituent-quark mass employed in usual quark models, i.e. $3M_f\approx M_\mathrm{nucleon}$. Applying this value of $M_{f}$ into the second equation, we have $\langle iq^{\dagger}q\rangle \equiv-\langle \bar{q}q\rangle \approx(235\,\mathrm{MeV})^{3}$ for the SU(2) light-flavor sector. Again, this value of the chiral condensate is well matched with phenomenologically accepted ones.

In contrast to these seemingly successful numerical results for the chiral limit, the LO results for the $M_{f}$ with finite $m_{f}$ is considerably deviated from the available LQCD data~\cite{Nam:2009jb,Nam:2009hq,Nam:2010nk}. In Refs.~\cite{Goeke:2007bj,Nam:2008bq}, it was suggested that the correct $m_f$ dependence of $M_{f}$ can only be achieved by the inclusion of the meson-loop corrections (MLC) which is related to the next-to-leading order (NLO) of the large-$N_{c}$ corrections. It has been also discussed that this NLO contributions play a critical role to reproduce the appropriate universal-class pattern of the chiral restoration as a function of $T$~\cite{Nam:2010nk}. Consequently we will discuss the inclusion of the MLC and the magnetic field in the next Section.
\section{E$\chi$A with MLC and $\bm{B}$ field}
Here we use a standard functional method~\cite{Goeke:2007bj,Nam:2008bq}
to tackle the MLC corresponding to the large-$N_{c}$ corrections.
Taking into account the mesonic fluctuations around their saddle-point values, one can write the E$\chi$A via a standard functional method as follows:
\begin{eqnarray}
\label{eq:EA2}
\mathcal{S}_{\mathrm{eff}}[m_{f}]&=&\mathcal{C}
+\mathcal{N}\ln\lambda+2\sigma^{2}
-\underbrace{\int\frac{d^4q}{(2\pi)^4}\mathrm{Tr}_{c,f,\gamma}
\ln\left[\frac{\rlap{/}{k}_a+i\bar{M}_{a} }{\rlap{/}{k}_a
+im_{f}}\right]}_\mathrm{LO}
\cr
&+&\underbrace{\frac{1}{2}\sum_{i=1}^{4}\int\frac{d^4k}{(2\pi)^4}
\ln\left\{1-\frac{1}{4\sigma^{2}}
\int\frac{d^4k}{(2\pi)^4}\mathrm{Tr}_{c,f,\gamma}
\left[ \frac{M_{a}}{\rlap{/}{k}_a+i\bar{M}_{a}}\Gamma_{i}
\frac{M_{b}}{\rlap{/}{k}_b+i\bar{M}_{b}}
\Gamma_{i}\right]\right\}}_\mathrm{NLO},
\end{eqnarray}
where $\Gamma_{i}=(1,\gamma_{5},i\bm{\tau},i\bm{\tau}\gamma_{5})$ relates to the fluctuations from the isoscalar-scalar, isoscalar-pseudoscalar, isovector-scalar, and isovector-pseudoscalar mesons with the Pauli matrix denoted by $\bm{\tau}$. As for the saddle-point values, we integrated out all the meson contributions except for the scalar one which signals for the SB$\chi$S.  For more details for Eq.~(\ref{eq:EA2}) can be found in Refs.~\cite{Goeke:2007bj,Nam:2008bq}. The $k_a$ and $k_b$ denote $k$ and $k+q$, respectively.

To study the impact of the external EM field on the QCD matter we need embed the external EM field in the E$\chi$A.
Since we are only interested in the external magnetic field, assuming that it is static and aligned along the $z$ axis as $\bm{B}=B_{0}\hat{z}$, we can choose the EM field configuration as follows:
\begin{equation}
\label{eq:AAA}
A_{\mu}=\left(-\frac{B_{0}}{2}y,\frac{B_{0}}{2}x,0,0 \right).
\end{equation}
This EM field configuration in Eq.~(\ref{eq:AAA}) makes the field-strength tensor, satisfying $B_{0}=F_{12}$. Employing the linear Schwinger method~\cite{Schwinger:1951nm,Nam:2009jb,Nam:2009hq}, one can write the E$\chi$A in Eq.~(\ref{eq:EA2}) as a function of the EM field strength:
\begin{eqnarray}
\label{eq:EA3}
\mathcal{S}_{\mathrm{eff}}[m_{f},F_{\mu\nu}]&\approx&\mathcal{C}
+\mathcal{N}\ln\lambda+2\sigma^{2}_{F}
-\int\frac{d^4k}{(2\pi)^4}\mathrm{Tr}_{c,f,\gamma}
\ln\left[\frac{\rlap{\,/}{K}_a+i\bar{M}_{A} }
{\rlap{\,/}{K}_a+im_{f}}\right]
\cr
&+&\frac{1}{2}\sum_{i=1}^{4}\int\frac{d^4q}{(2\pi)^4}
\ln\left\{1-\frac{1}{4\sigma^{2}_{F}}
\int\frac{d^4k}{(2\pi)^4}\mathrm{Tr}_{c,f,\gamma}
\left[ \frac{M_{a}}{\rlap{/}{k}_a+i\bar{M}_{a}}\Gamma_{i}
\frac{M_{b}}{\rlap{/}{k}_b+i\bar{M}_{b}}
\Gamma_{i}\right]\right\}.
\end{eqnarray}
Here, the subscript $F$ denotes the quantities under the external EM field. In Eq.~(\ref{eq:EA3}), we use the notation $K_{\mu}=k_{\mu}+e_{f}A_{\mu}$. Here $e_f=Q_f e$, here $e$ is the electric charge of the proton.
Note that we have assumed that the NLO part is not modified by the external EM field, since the EM contribution from the NLO is small in comparison to the leading one. In the presence of the external EM field, the quark propagator is modified and approximated as~\cite{Nam:2009jb,Nam:2009hq,Nam:2008ff}:
\begin{equation}
\label{eq:BPRO}
\frac{1}{\rlap{\,/}{K}_a+i\bar{M}_{a}}
\approx\frac{\rlap{\,/}{K}_a-i[\bar{M}_{a}+e_{f}N_{a}(\sigma\cdot F)]}
{k^{2}_a+\bar{M}^{2}_{a}}+\mathcal{O}(Q^{n\ge3}_f),
\end{equation}
where we have ignored the terms proportional to $\mathcal{O}(Q^{n\ge3}_f)$, taking into account that $(Q_u,Q_d)=(2/3,-1/3)$. The relevant mass functions are also defined by the following expressions:
\begin{equation}
\label{eq:MMMMM}
\bar{M}_{a}=m_{f}+M_{a}
=m_{f}+M_{f}\left(\frac{2}{2+k^{2}_a\bar{\rho}^{2}} \right)^{2},
\,\,\,\,
N_{a}=-\frac{4M_{f}\bar{\rho}^{2}}
{(2+k^{2}_a\bar{\rho}^{2})^{3}}.
\end{equation}
From Eq.~(\ref{eq:EA3}), the LO contributions for the instanton number density $\mathcal{N}$ in  Eq.~(\ref{eq:SCE}) and chiral condensate in Eq.~(\ref{eq:CC}), in the presence of the external magnetic field, can be derived up to $\mathcal{O}(Q^{2}_{f})$:
\begin{eqnarray}
\label{eq:LO}
\mathcal{N}_{\mathrm{LO},F}
&=&2N_{c}N_{f}\int\frac{d^4k}{(2\pi)^4}
\left[\frac{M_{a}\bar{M}_{a}}
{k^{2}_a+\bar{M}^{2}_{a}}
+\frac{2N^{2}_{a}\mathcal{B}^{2}_{f}}
{k^{2}_a+\bar{M}^{2}_{a}}\right],
\cr
\langle iq^{\dagger}q\rangle_{\mathrm{LO},F}&=&
4N_{c}\int\frac{d^4k}{(2\pi)^4}
\left[\frac{\bar{M}_{a}}{k^{2}_a+\bar{M}^{2}_{a}}
-\frac{m_{f}}{k^{2}_a+m^{2}_{f}}\right],
\end{eqnarray}
where we assigned $\mathcal{B}_{f}$ as $e_{f}B_{0}=Q_{f}(eB_{0})$ for simplicity, and the terms proportional to $\mathcal{O}(Q_{f})$ does not appear in the $\mathcal{N}_\mathrm{LO,F}$, due to $\mathrm{Tr}_{\gamma}(\sigma\cdot F)=0$. If $B_{0}=0$ the expressions for the $\mathcal{N}$ and chiral condensate in Eq.~(\ref{eq:LO}) recover those for vacuum given in Section II. Note that, although we do not have the explicit terms $\propto\mathcal{B}^2$ for the chiral condensate as far as we employ the quark propagator in Eq.~(\ref{eq:BPRO}), the condensate depends on the magnetic field because the $M_f$ itself does implicitly. Moreover, the constituent-quark mass under the external magnetic field, which is determined from the saddle-point equation in the first line of Eq.~(\ref{eq:LO}), becomes different for the two flavors $u$ and $d$, i.e. $M_{u,F}\ne M_{d,F}$. It is because that they behave distinctively according to their electric charges. Similarly the chiral condensate also becomes flavor-dependent as shown in the second line of Eq.~(\ref{eq:LO}). Here is one caveat: For all the ingredients discussed so far, we have assumed that the instanton-packing fraction, $\mathcal{N}$ is immune from the external magnetic field as well as the flavor degrees of freedom by considering that the (anti)instanton is electrically neutral and non-flavored object.

It is worth mentioning the differences between our theoretical framework and other chiral models. For instance, using the NJL model, Refs.~\cite{Menezes:2008qt,Boomsma:2009yk} obtained the magnetic-field dependent effective action. Most apparent difference between the present approach and theirs is how to regularize the UV divergence appearing in relevant physical quantities. The UV divergence is regularized by the nonlocal quark-instanton interaction in the present approach (see the quark form factors given in Eqs.~(\ref{eq:MDEQM}) and  (\ref{eq:FFPARA})), whereas the regularization is achieved by adding and subtracting the lowest-Landau level (LLL) and the vacuum contribution of the chiral order parameters in the NJL model~\cite{Menezes:2008qt,Boomsma:2009yk}. In this sense, in our approach, all the Landau levels are taken into account by construction in principle. Hence we do not need any specific analytic manipulations.  A more detailed discussions on this regularization process based on the Landau level for the chiral models is given in Ref.~\cite{Gatto:2010pt} where the highest levels are naturally cut off, since the constituent mass at large momenta is suppressed when the magnetic becomes strong.

The numerical value for $\sigma_{F}$ in Eq.~(\ref{eq:LO}) is obtained from the relation $\sigma^{2}=\mathcal{N}/2$ in the LO contributions as in Ref.~\cite{Nam:2009nn}. Thus, using the LO part in the right-hand side of Eq.~(\ref{eq:LO}), we can write as follows:
\begin{equation}
\label{eq:SIGMA}
\sigma^{2}_{F}=\frac{\mathcal{N}_{\mathrm{LO},F}}{2}
+\mathrm{NLO\,contributions}\approx\frac{\mathcal{N}_{\mathrm{LO},F}}{2},
\end{equation}
where we have rather safely ignored the NLO ones for the numerical calculations, since the NLO contributions are finite but much small in comparison to the LO one. By doing this, one can express the $\sigma^{2}_{F}$ simply as a function of the external magnetic field. Considering all the ingredients discussed so far, we write the $\mathcal{N}$ containing  both of the LO and  NLO (MLC) contributions, using Eq.~(\ref{eq:SIGMA}):
\begin{eqnarray}
\label{eq:NOVMLC}
\mathcal{N}&\approx&
2N_{c}N_{f}\int\frac{d^4k}{(2\pi)^4}[F_1(k)+F_2(k)\mathcal{B}^{2}_{f}]
+\frac{3}{2}\frac{\int\frac{d^4q}{(2\pi)^4}\frac{d^4k}{(2\pi)^4}\,F_3(k,q)}
{\int\frac{d^4k}{(2\pi)^4}[F_1(k)+F_2(k)\mathcal{B}^{2}_{f}]},
\end{eqnarray}
where the reduced functions $F_{1\sim3}$ are defined as
\begin{eqnarray}
\label{eq:FFFFF}
F_1(k)&=&\frac{M_a\bar{M}_a}{D^2_a},\,\,\,\,F_2(k)=\frac{2N^2_a}{D^2_a},\,\,\,\,
F_3(k,q)=\frac{M_aM_b
[k_a\cdot k_b+\bar{M}_{a}\bar{M}_{b}+M_{a}M_{b}+\frac{m_f}{2}(M_{a}+M_{b})]}{D^2_aD^2_b}.
\end{eqnarray}
Here, we have used the notations $D^2_{a,b}=k^2_{a,b}+\bar{M}^2_{a,b}$. Similarly, the chiral condensate can be evaluated from Eq.~(\ref{eq:CC}) as follows:
\begin{eqnarray}
\label{eq:CCMLC}
\langle iq^{\dagger}q\rangle_{\mathrm{LO+NLO},F}\approx
4N_{c}\int\frac{d^4k}{(2\pi)^4}[G_1(k)+G_2(k)]+\frac{3}{2N_f}\frac{\int\frac{d^4q}
{(2\pi)^4}\frac{d^4k}{(2\pi)^4}\,G_3(k,q)}{\int\frac{d^4k}{(2\pi)^4}
[F_1(k)+F_2(k)\mathcal{B}^2_f]}.
\end{eqnarray}
Here, the functions $G_{1\sim3}$ are assigned as
\begin{eqnarray}
\label{eq:GGGG}
G_1(k)&=&\frac{\bar{M}_a}{D^2_a},\,\,\,\,G_2(k)=-\frac{m_f}{D^2_0},\,
G_3(k,q)=\frac{M_aM_b(\bar{M}_a+\bar{M}_b)}{D^2_aD^2_b},
\end{eqnarray}
where $D^2_0=k^2+m^2_f$. It is interesting to see from Eqs.~(\ref{eq:LO}) and (\ref{eq:CCMLC}) that, if the isospin symmetry is almost intact, $m_{u}\approx m_{d}$, the difference between the condensates of the two flavors, i.e $\langle iu^{\dagger}u\rangle-\langle id^{\dagger}d\rangle$, becomes negligible for the case with $\mathcal{B}=0$. However, as the strength of the magnetic field increases, the difference is also proportional to $(e^{2}_{u}-e^{2}_{d})B^{2}_{0}$. For a better look on the isospin breaking effect, we define a quantity indicating the strength of the isospin breaking effect as follows:
\begin{equation}
\label{eq:RATIO}
\mathcal{R}\equiv
\frac{\langle iu^{\dagger}u\rangle-\langle id^{\dagger}d\rangle}
{\langle iu^{\dagger}u\rangle+\langle id^{\dagger}d\rangle}.
\end{equation}
We also note that the ratio $\mathcal{R}$ is deeply related to the low-energy constant of the $\chi$PT Lagrangian, $h_{3}$~\cite{Gasser:1983yg,Goeke:2010hm}.
\section{Instanton parameters at finite $T$}
To investigate the physical quantities in hand at finite $T$, we want to discuss briefly how to modify the instanton parameters, $\bar{\rho}$ and $\bar{R}$ at finite $T$. We will follow our previous work~\cite{Nam:2009nn} and Refs.~\cite{Harrington:1976dj,Diakonov:1988my} to this end. Usually, there are two different instanton configurations at finite $T$, being periodic in Euclidean time, with trivial and nontrivial holonomies. They are called the Harrington-Shepard~\cite{Harrington:1976dj} and Kraan-van Baal-Lee-Lu calorons~\cite{Kraan:1998pm,Lee:1998bb}, respectively. The nontrivial holonomy can be identified as the Polyakov line as an order parameter for the confinement-deconfinement transition of QCD. However, since we are not interested in the confinement-deconfinement transition in the present work, we choose the Harrington-Shepard caloron for the parameter modifications at finite $T$. We write the instanton distribution function at finite $T$ with the Harrington-Shepard caloron as follows:
\begin{equation}
\label{eq:IND}
d(\rho,T)=\underbrace{C_{N_c}\,\Lambda^b_{\mathrm{RS}}\,
\hat{\beta}^{N_c}}_\mathcal{C}\,\rho^{b-5}
\exp\left[-(A_{N_c}T^2
+\bar{\beta}\gamma{\cal N}\bar{\rho}^2)\rho^2 \right].
\end{equation}
Here, the abbreviated notations are also given as:
\begin{equation}
\label{eq:para}
\hat{\beta}=-b\ln[\Lambda_\mathrm{RS}\rho_\mathrm{cut}],\,\,\,\,
\bar{\beta}=-b\ln[\Lambda_\mathrm{RS}\langle R\rangle],\,\,\,
C_{N_c}=\frac{4.60\,e^{-1.68\alpha_{\mathrm{RS}} Nc}}{\pi^2(N_c-2)!(N_c-1)!},
\end{equation}
\begin{equation}
\label{eq:AA}
A_{N_c}=\frac{1}{3}\left[\frac{11}{6}N_c-1\right]\pi^2,\,\,\,\,
\gamma=\frac{27}{4}\left[\frac{N_c}{N^2_c-1}\right]\pi^2,\,\,\,\,
b=\frac{11N_c-2N_f}{3},\,\,\,\,{\cal N}=\frac{N}{V}.
\end{equation}
Note that we defined the one-loop inverse charge $\hat{\beta}$ and $\bar{\beta}$ at a certain phenomenological cutoff value $\rho_\mathrm{cut}$ and $\langle R\rangle\approx\bar{R}$. As will be shown, only $\bar{\beta}$ is relevant in the following discussions and will be fixed self-consistently within the present framework. $\Lambda_{\mathrm{RS}}$ stands for a scale depending on a renormalization scheme, whereas $V_3$ stands for the three-dimensional volume. Using the instanton distribution function in Eq.~(\ref{eq:IND}), we can compute the average value of the instanton size, $\bar{\rho}^2$ straightforwardly as follows~\cite{Schafer:1996wv}:
\begin{equation}
\label{eq:rho}
\bar{\rho}^2(T)
=\frac{\int d\rho\,\rho^2 d(\rho,T)}{\int d\rho\,d(\rho,T)}
=\frac{\left[A^2_{N_c}T^4
+4\nu\bar{\beta}\gamma {\cal N}\right]^{\frac{1}{2}}
-A_{N_c}T^2}{2\bar{\beta}\gamma {\cal N}},
\end{equation}
where $\nu=(b-4)/2$. Substituting Eq.~(\ref{eq:rho}) into Eq.~(\ref{eq:IND}), the distribution function can be evaluated further as:
\begin{equation}
\label{eq:dT}
d(\rho,T)=\mathcal{C}\,\rho^{b-5}
\exp\left[-\mathcal{M}(T)\rho^2 \right],\,\,\,\,
\mathcal{M}(T)=\frac{1}{2}A_{N_c}T^2+\left[\frac{1}{4}A^2_{N_c}T^4
+\nu\bar{\beta}\gamma {\cal N}\right]^{\frac{1}{2}}.
\end{equation}
The instanton-number density ${\cal N}$ can be computed self-consistently as a function of $T$, using the following equation:
\begin{equation}
\label{eq:NOVV}
{\cal N}^\frac{1}{\nu}\mathcal{M}(T)=\left[\mathcal{C}\,\Gamma(\nu) \right]^\frac{1}{\nu},
\end{equation}
where we have replaced $NT/V_3\to {\cal N}$, and $\Gamma(\nu)$ indicates a $\Gamma$ function with an argument $\nu$. Note that $\mathcal{C}$ and $\bar{\beta}$ can be determined easily using Eqs.~(\ref{eq:rho}) and (\ref{eq:NOVV}), incorporating the vacuum values of the ${\cal N}$ and $\bar{\rho}$: $\mathcal{C}\approx9.81\times10^{-4}$ and $\bar{\beta}\approx9.19$. At the same time, using these results, we can obtain the average instanton size $\bar{\rho}$ as a function of $T$ with Eq.~(\ref{eq:rho}).

\begin{figure}[t]
\begin{tabular}{cc}
\includegraphics[width=8.5cm]{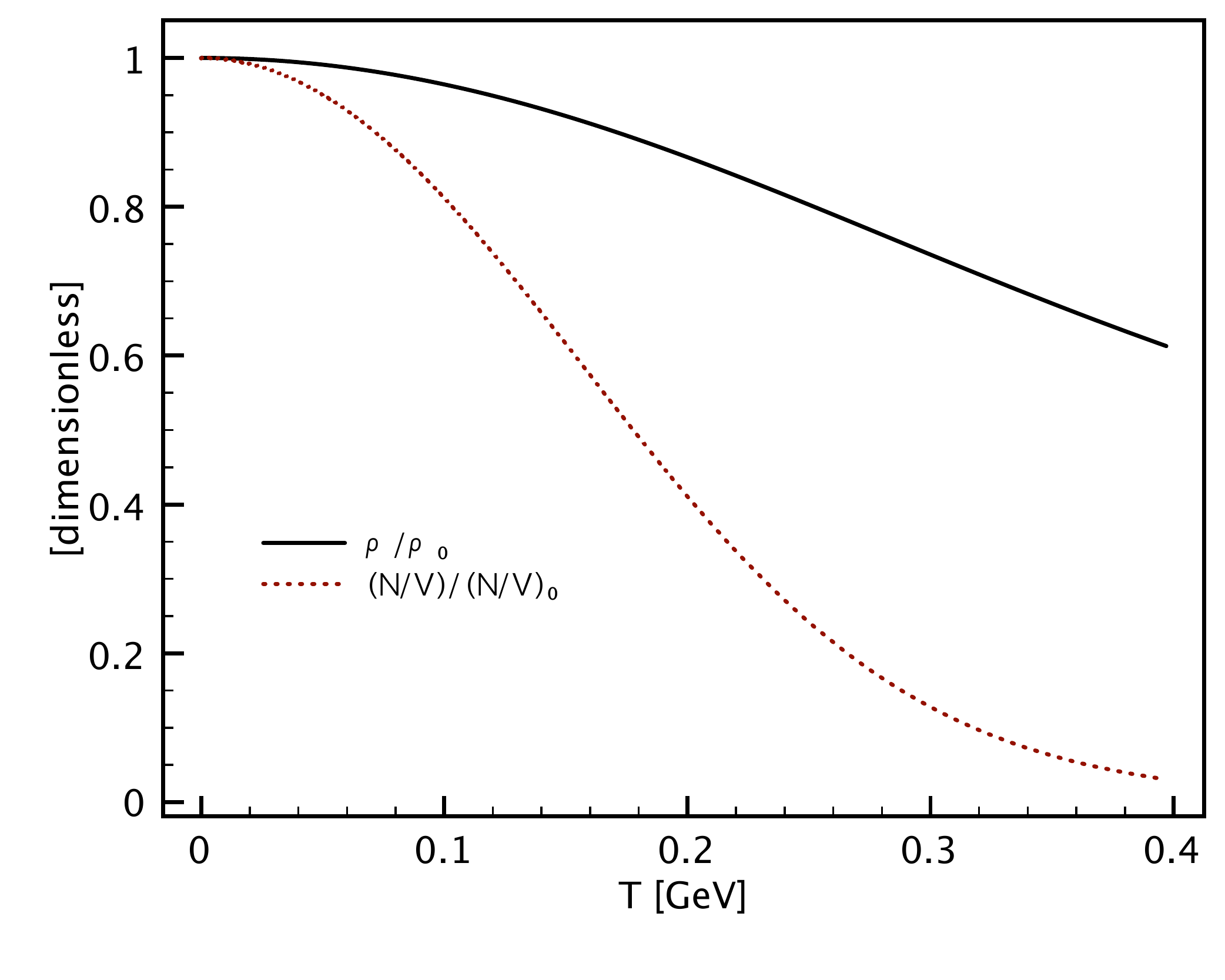}
\includegraphics[width=9.5cm]{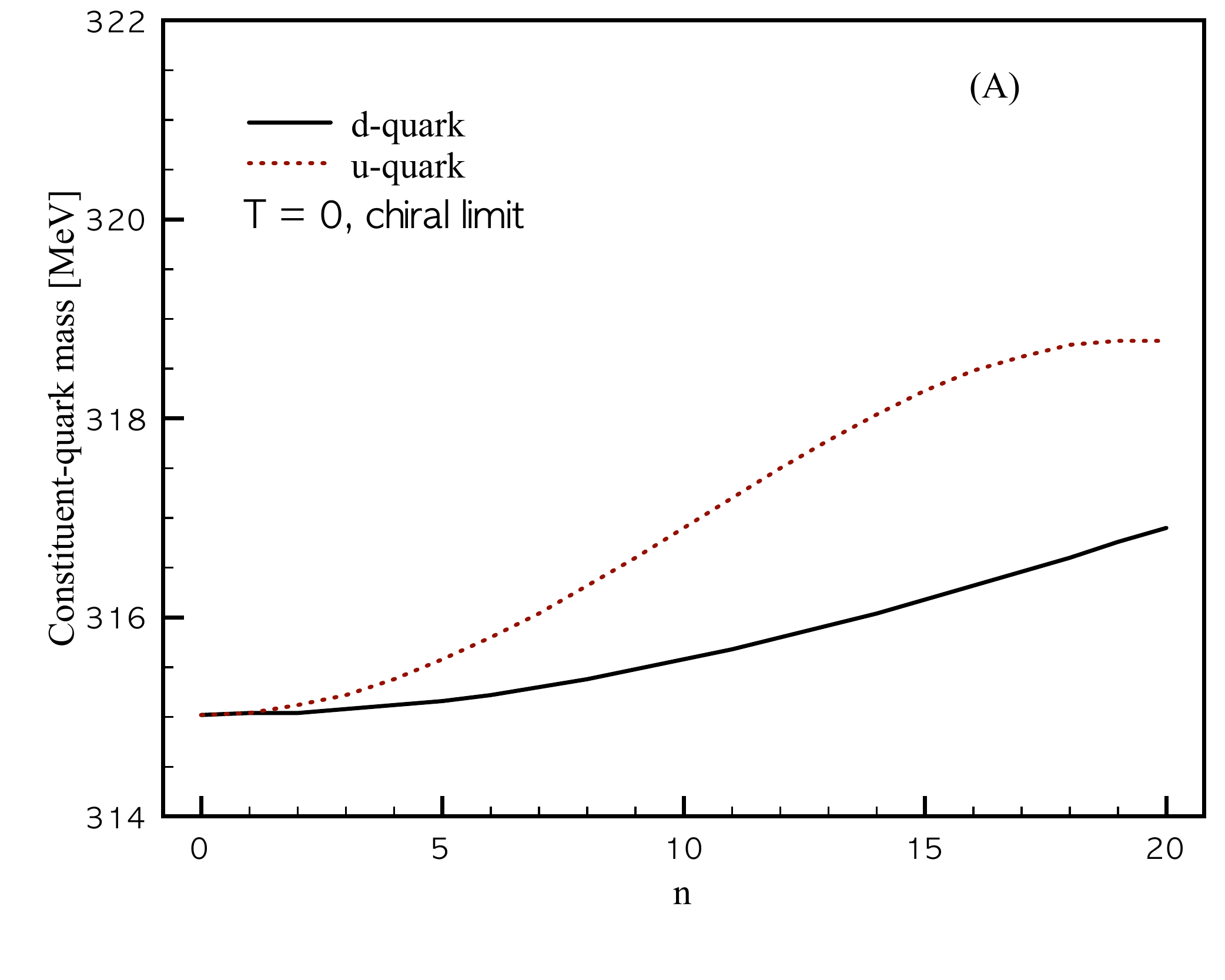}
\end{tabular}
\caption{(Color online) Normalized $\bar{\rho}/\bar{\rho}_{0}$ and ${\cal N}/{\cal N}_{0}$ as a function of $T$ for $N_{c}=3$, where $\mathcal{N}\equiv N/V$ (left). The $M(\bm{k}^2,T)$ in Eq.~(\ref{eq:M00}) as a function of $T$ and absolute value of the momentum $|\bm{k}|$ (right).}
\label{FIG1}
\end{figure}

Finally, in order to estimate the $T$ dependence of the constituent-quark mass $M_{f}$, it is necessary to consider the normalized distribution function, defined as follows:
\begin{equation}
\label{eq:NID}
d_N(\rho,T)=\frac{d(\rho,T)}{\int d\rho\,d(\rho,T)}
=\frac{\rho^{b-5}\mathcal{M}^\nu(T)
\exp\left[-\mathcal{M}(T)\rho^2 \right]}{\Gamma(\nu)}.
\end{equation}
Now, we want to employ the large-$N_c$ limit to simplify the expression of $d_N(\rho,T)$. Since the parameter $b$ is in the order of $\mathcal{O}(N_c)$ as shown in Eq.~(\ref{eq:para}), it becomes infinity as $N_c\to\infty$, and the same is true for $\nu$. In this limit, as understood from Eq.~(\ref{eq:NID}), $d_N(\rho,T)$ can be approximated as a $\delta$ function~\cite{Diakonov:1995qy}:
\begin{equation}
\label{eq:NID2}
\lim_{N_c\to\infty}d_N(\rho,T)=\delta({\rho-\bar{\rho}}).
\end{equation}
Remember that the constituent-quark mass can be represented by~\cite{Diakonov:1995qy}
\begin{equation}
\label{eq:M0}
M_{f}\propto\sqrt{\mathcal{N}}
\int d\rho\,\rho^{2}\delta(\rho-\bar{\rho})
=\sqrt{\mathcal{N}}\,\bar{\rho}^{2},
\end{equation}
where $\mathcal{N}$ and $\bar{\rho}$ are functions of $T$ implicitly. We can modify $M_{f}$ as a function of $T$ as follows:
\begin{equation}
\label{eq:momo}
M_{f}\to M_{f}\left(\frac{\sqrt{\mathcal{N}}}
{\sqrt{\mathcal{N}_{0}}}
\frac{\bar{\rho}^2}{\bar{\rho}^2_{0}}\right)\equiv M_{f}(T)
\end{equation}
where $\mathcal{N}_{0}$ and $\bar{\rho}_{0}$ are those at $T=0$. The numerical results for the normalized $\bar{\rho}/\bar{\rho}_{0}$ and ${\cal N}/{\cal N}_{0}$ as functions of $T$ are in the left panel of Figure~\ref{FIG1}. As shown there, these quantities are decreasing with respect to $T$ as expected: decreasing instanton effect. However, even beyond $T^{\chi}_{c}\approx\Lambda_{\mathrm{QCD}}\approx200$ MeV, the instanton contribution remains finite. In the right panel of figure, we draw the quark mass as a function of $T$ and absolute value of three momentum of a quark $|\bm{k}|$:
\begin{equation}
\label{eq:M00}
M(\bm{k}^2,T)=M_{f}(T)\left[\frac{2}{2+\bar{\rho}^{2}(T)\bm{k}^{2}}\right].
\end{equation}
Note that we have ignored the Euclidean-time component of the four momentum by setting $k_{4}=0$. This tricky treatment simplifies the calculations in hand to a large extent, and we also verified that only a small deviation appears in comparison to the full calculations. Moreover, $\bar{\rho}$ in Eq.~(\ref{eq:M00}) is now a function of $T$ as demonstrated by Eqs.~(\ref{eq:rho}) and (\ref{eq:momo}) previously. As shown in the figure, $M(|\bm{k}|,T)$ is a smoothly decreasing function of $T$ and $|\bm{k}|$, indicating that the effect of the instanton is diminished.  Here, we choose $M_f=325$ MeV at $T=0$ in drawing the curve as a trial. For more details, one can refer to the previous work~\cite{Nam:2009nn}.

\section{E$\chi$A with MLC and external magnetic field at finite $T$}
In this section we generalize our model to the finite temperature to explore the chiral restoration at finite $T$. For this purpose, we employ the Matsubara formula for the fermions. In this Euclidean-time description, it can be done by replacing the four-dimensional integral  measure in the E$\chi$A into a three-dimensional one with a summation over the fermionic Matsubara frequency, $w_{m}=(2m+1)\pi T$:
\begin{equation}
\label{eq:MATS}
\int\frac{d^4k}{(2\pi)^4}\to
T\sum^{\infty}_{m=-\infty}\int\frac{d^3\bm{k}}{(2\pi)^3}.
\end{equation}
Using this simple replacement we can rewrite the saddle-point equation in Eq.~(\ref{eq:NOVMLC}) as follows:
\begin{eqnarray}
\label{eq:NOVT}
\mathcal{N}&\approx&
2N_{c}N_{f}\int\frac{d^3\bm{k}}{(2\pi)^3}
[\mathcal{F}_1(\bm{k})+\mathcal{F}_2(\bm{k})\mathcal{B}^{2}_{f}]
+\frac{3\Lambda}{4\pi}\frac{\int\frac{d^3\bm{q}}{(2\pi)^3}\frac{d^3\bm{k}}{(2\pi)^3}
[\mathcal{F}_3(\bm{k},\bm{q})+\mathcal{F}_4(\bm{k},\bm{q})]}
{\int\frac{d^3\bm{k}}{(2\pi)^3}[\mathcal{F}_1(\bm{k})+\mathcal{F}_2(\bm{k})\mathcal{B}^{2}_{f}]}.
\end{eqnarray}
Summing over the Matsubara frequency, we define functions $\mathcal{F}_i$ as follows:
\begin{equation}
\label{eq:}
\mathcal{F}_i\equiv T\sum_{m} F_i.
\end{equation}
The analytic expressions for those functions are given in the Appendix. Note that the above equation recovers the expression of Eq.~(16) in Ref.~\cite{Nam:2010mh} when $\mathcal{B}=0$. Before going further, it is worth mentioning several assumptions made for deriving Eq.~(\ref{eq:NOVT}):
\begin{itemize}
\item The instanton packing fraction, $\mathcal{N}$ is a decreasing function of $T$ indicated by the previous Section. However, we assume that the $\mathcal{N}$ is not affected by the external magnetic field, since the (anti)instantons are electrically neutral. Moreover, the immunity of $\mathcal{N}$ to flavors are assumed.
\item In the momentum-dependent quark mass $M_{a,b}$, we replace $k^{2}$ into $\bm{k}^{2}$, ignoring the temporal term $\propto k_{4}=(2m+1)\pi T$. We verified that this treatment makes the numerical calculations much simpler, and only small deviation was observed in comparison to the full calculations. Hence, the mass-related functions in Eqs.~(\ref{eq:FFPARA}) and (\ref{eq:MMMMM}) are redefined as follows:
\begin{equation}
\label{eq:MAAA}
M_{a}=\frac{4M_{f}}
{(2+\bm{k}^{2}\bar{\rho}^{2})^{2}},
\,\,\,\,
N_{a}=-\frac{4M_{f}\bar{\rho}^{2}}
{(2+\bm{k}^{2}\bar{\rho}^{2})^{3}}.
\end{equation}
As for the $M_{b}$ and $N_{b}$, we replace $\bm{k}$ with $\bm{k}+\bm{q}$.
\item Similarly, the term $k_a\cdot k_b=k^2+k\cdot q$ in the functions of $F_i$ and $G_i$ is replaced by $w^2_{m}+\bm{k}^2+\bm{k}\cdot\bm{q}$. Moreover, the denominator is also replaced by $D^2_{a,b}=w^2_{m}+\bm{k}^2_{a,b}+\bar{M}^2_{a,b}$.
\item We also replace the integral variable $q_{4}$, which corresponds to the fourth-component of the pion momentum, into an additional parameter $\Lambda$ as in Ref.~\cite{Nam:2008bq}. Since the isovector-pseudoscalar meson, i.e. pion dominates the meson fluctuations, it is reasonable to set the cutoff $\Lambda$ proportional to $m_{\pi}$ as follows:
\begin{equation}
\label{eq:LAMBDA}
\Lambda\approx m_{\pi}\frac{\bar{\rho}_{0}}{\bar{\rho}}.
\end{equation}
Note that, in the above equation, we have multiplied a factor $\bar{\rho}_{0}/\bar{\rho}$ to $m_{\pi}$ in order to include $T$ dependence of the cutoff mass. Moreover, this multiplication factor represents a correct chiral-restoration pattern of $m_{\pi}$, i.e. the mass of the pion, as a Nambu-Goldstone (NG) boson, increases as SB$\chi$S restored partially.
\end{itemize}

Similarly the chiral condensate in Eq.~(\ref{eq:CCMLC}) reads:
\begin{eqnarray}
\label{eq:CCT}
\langle iq^{\dagger}q\rangle\approx
4N_{c}\int\frac{d^3\bm{k}}{(2\pi)^3}
\left[\mathcal{G}_1(\bm{k})+\mathcal{G}_2(\bm{k})\right]+\frac{3\Lambda}{4\pi N_f}
\frac{\int\frac{d^3\bm{q}}{(2\pi)^3}\frac{d^3\bm{k}}{(2\pi)^3}\,
\mathcal{G}_3(\bm{k},\bm{q})}
{\int\frac{d^3\bm{k}}{(2\pi)^3}[\mathcal{F}_1(\bm{k})+\mathcal{F}_2(\bm{k})
\mathcal{B}^2_f]}.
\end{eqnarray}
Again, the relevant functions $\mathcal{G}_i\equiv T\sum_{m}G_{i}$ are defined and given in the Appendix.

\section{Numerical results}
We present and discuss our numerical results in this Section. We choose $\bar{R}\approx1.0$ fm and $\bar{\rho}\approx0.34$ fm which give $M_f\approx315$ MeV in the present framework~\cite{Goeke:2010hm}. The values for the current-quark mass are reported as $m_{u}=(1.7\sim3.3)$ MeV and $m_{d}=(4.1\sim5.8)$ MeV~\cite{Nakamura:2010zzi}.  Thus, we take the average values, $(m_{u},m_{d})\approx(2.5,5)$ MeV. For simplicity we introduce an positive integer $n$. The magnetic field is assigned in terms of the pion mass as $eB_{0}=n\,m^{2}_{\pi}$. In the Gauss unit for the magnetic field, we have the convention, $B_0\approx n\,(1.2\times10^{18})\,\mathrm{G}$. For instance, $n\approx1$ corresponds to a neutron star or magnetar with very strong magnetic field. The case with $n\approx10 $ or more may be observed inside the quark-gluon plasm created in the ultra high-energy peripheral heavy-ion collisions, such as RHIC and LHC, as the main source for the nontrivial QCD vacuum effect, i.e. chiral magnetic effect~\cite{Kharzeev:2004ey,Voloshin:2004vk,Kharzeev:2007jp,Voloshin:2008jx,:2009txa,:2009uh,Fukushima:2008xe,Fukushima:2009ft,Warringa:2009rw,Asakawa:2010bu}.

\subsection{Constituent-quark mass for each flavor: $M_f$}
Here we present the numerical result of the $T$ and $B_{0}$ dependencies of the constituent-quark mass $M_f$ which is one of the order parameters for the (partial) chiral restoration. According to the universal class of the restoration pattern, as for the chiral limit $m_{u,d}=0$, the restoration  pattern shows the second order, whereas it becomes the crossover for the case with the physical quark mass, i.e. $m_{u,d}=(2.5,5)$ MeV.  In our previous work~\cite{Nam:2010mh}, the MLC contributions, as the large-$N_c$ corrections, play a critical role to reproduce the universality in an appropriate manner. Similar observation was also reported in Ref.~\cite{Muller:2010am}. Here we study the impact of the external magnetic field on
the two light-flavor QCD matter at finite $T$.

\begin{figure}[t]
\begin{tabular}{cc}
\includegraphics[width=8.5cm]{FIG2-1.pdf}
\includegraphics[width=8.5cm]{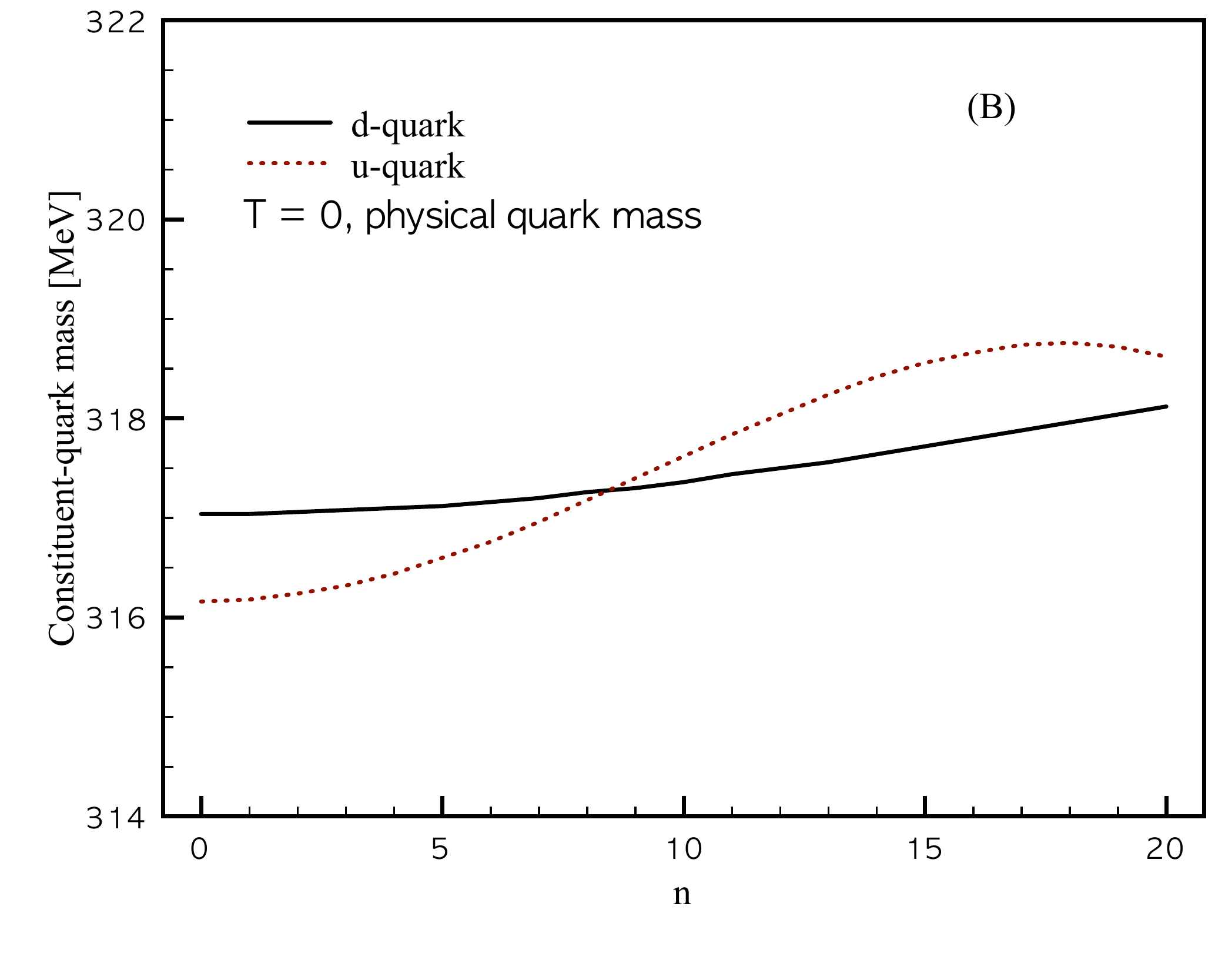}
\end{tabular}
\begin{tabular}{cc}
\includegraphics[width=8.5cm]{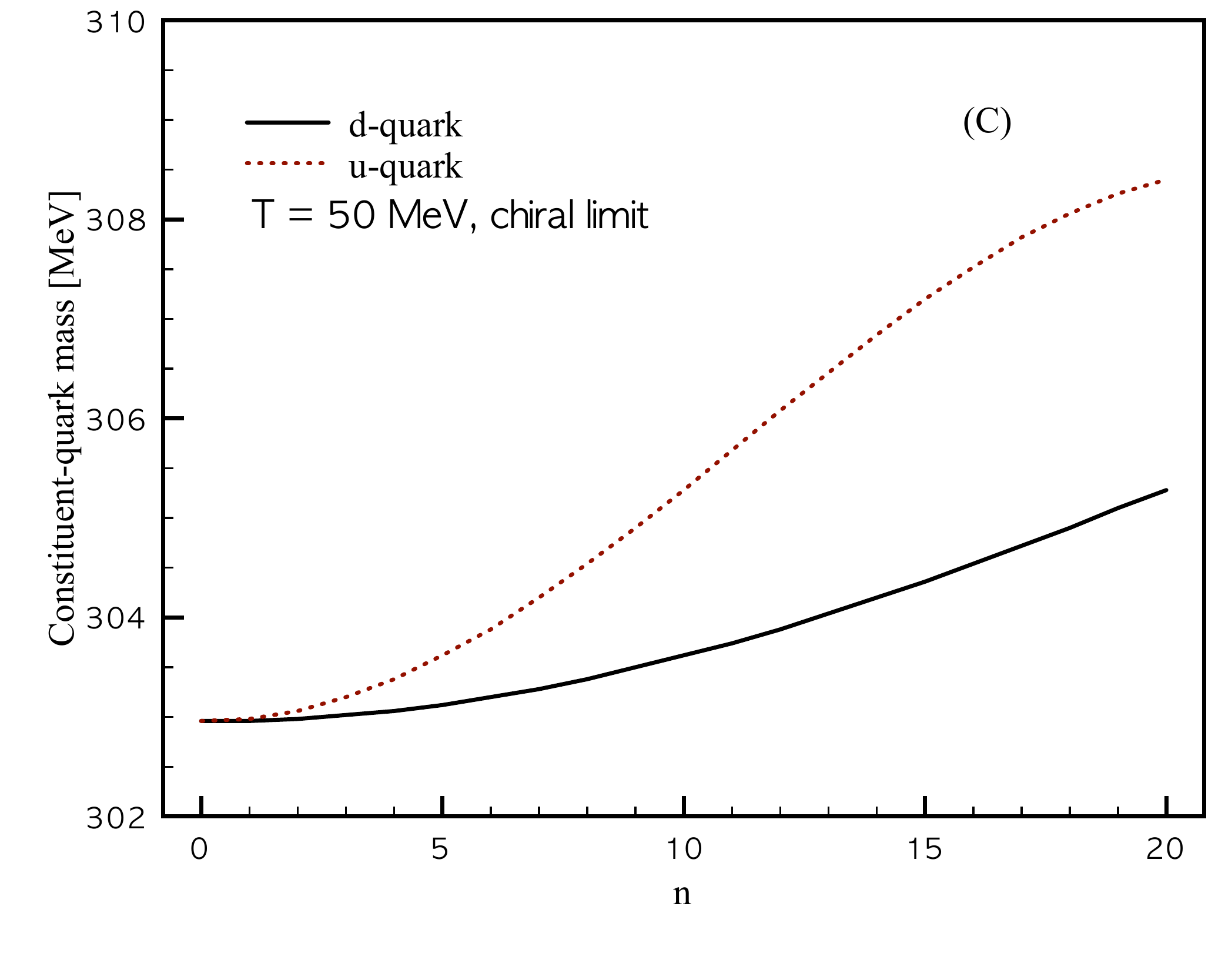}
\includegraphics[width=8.5cm]{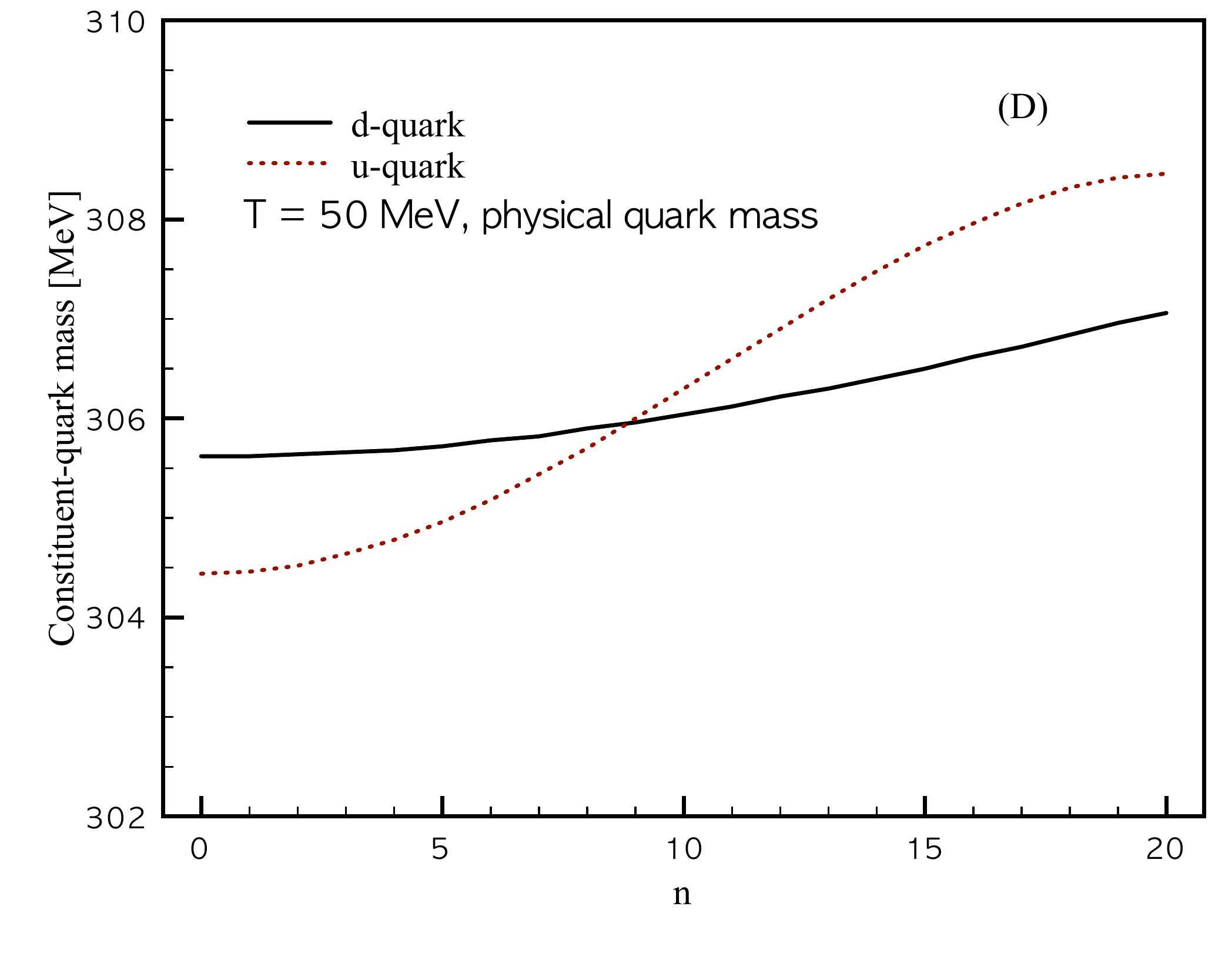}
\end{tabular}
\caption{Constituent-quark mass ($M_f$) as a function of $n\equiv eB_0/m^2_\pi$ for the chiral limit (left column) and physical quark mass (right column). In the upper and lower rows, we show the numerical results for $T=0$ and $T=50$ MeV, respectively. For more details, see the text.}
\label{FIG2}
\end{figure}

In Figure~\ref{FIG2}, we depict the results of the $M_f$ at $T=0$ for the chiral limit in the panel (A) and physical quark mass case in the panel (B). In the absence of the magnetic field $(n=0)$, we have $M_{u,d}=315.02$ MeV for the chiral limit, and $M_{u,d}=(316.16,317.04)$ MeV for the physical quark mass. Accounting for $m_u<m_d$ and that the constituent-quark mass becomes $M_{u,d}\sim M_f+m_f$, the observed results can be easily understood. As the external magnetic field emerges the constituent-quark mass becomes heavier. It is due to the magnetic catalysis. Since the effects of the magnetic catalysis is proportional to $e^2_f$ as in Eqs.~(\ref{eq:NOVT}) and (\ref{eq:CCT}), the $M_u$ grows rapidly more than the $M_d$ with respect to the external magnetic field, i.e. $e_u^2>e_d^2$.  Beyond $n=(17\sim18)$, the curve for the $M_u$ starts decreasing slightly. It is interesting to see that, for the physical quark mass case in the panel (B), there appears a point at which the $M_u$ and $M_d$ coincide each other ($n\approx$7.5). It is because that the broken isospin symmetry ($m_u\ne m_d$) is compensated by the effect from the external magnetic field. From our results, this interesting phenomena appears at the magnetic field $B_0\approx 10^{19}\,\mathrm{G}$, which can be created at the heavy-ion collision experiments. Beyond this point, the ordering of the $u$- and $d$-quark constituent masses are reversed.

Here, we want to discuss briefly the saturation behavior of the constituent-quark mass observed in the panel (A) and (B) beyond $n\approx15$ for the $M_u$. This can be understood as follows: As discussed in the previous Sections (see Eq.~(\ref{eq:BPRO}) for instance), we have taken into account the contributions up to $\mathcal{O}(Q_f^2)$. Hence, as for the stronger magnetic fields, higher-order contributions which are not included in the present work will be no longer negligible. Accordingly, we verified that this saturation behavior becomes weaker in the presence of higher-order contributions. Nevertheless. the complete and systematic treatment of those higher-order terms is left for the future work.

In the panel (C) and (D) the results for the case at $T=50$ MeV are demonstrated. Note that, for all the cases, the absolute values for the $M_f$ decrease in comparison to those at $T=0$. But the shapes and behaviors of the curves are similar. This decreasing tendency can be understood by the diluting instanton ensemble at finite $T$. For more details on the diluting ensemble in the present framework, one may refer to Refs.~\cite{Nam:2009nn}. It turns out that the decreasing rate of the constituent-quark mass is a few percent from $T=0$ to $T=50$ MeV: $M_{u,d}=(302.96)$ MeV for the chiral limit and $M_{u,d}=(304.44,305.62)$ MeV for physical quark mass, at $n=0$. Similarly to the vacuum case $T=0$, there appears a point $n\approx8$, at which the $u$- and $d$-quark constituent masses are very close to each other. Moreover, the isospin breaking effect $(m_d\ne m_u)$ becomes more obvious at $T=50$ MeV compared to that for $T=0$. It is because of the decreasing dynamically-generated quark mass at finite $T$. All of these observations are all based on a nontrivial competition between the magnetic catalysis and diluting instanton effects at finite $T$: The former tends to enhance the constituent quark mass, whereas the latter suppress it. Hence, the position of the equal mass point is a consequence of this nontrivial competition between the two mechanisms, on top of the explicit isospin symmetry breaking. The reversing order of mass  beyond the equal mass point is also observed at finite $T$. In Ref.~\cite{Boomsma:2009yk}, the authors found qualitatively the same results; monotonically increasing curves for the $M_f$ with respect to the external magnetic field was observed and the $M_u$ is more sensitive to it. However, the rate of increasing is much higher than ours.

\subsection{Chiral condensate: $\langle iq^\dagger q\rangle$}
\begin{figure}[t]
\begin{tabular}{cc}
\includegraphics[width=8.5cm]{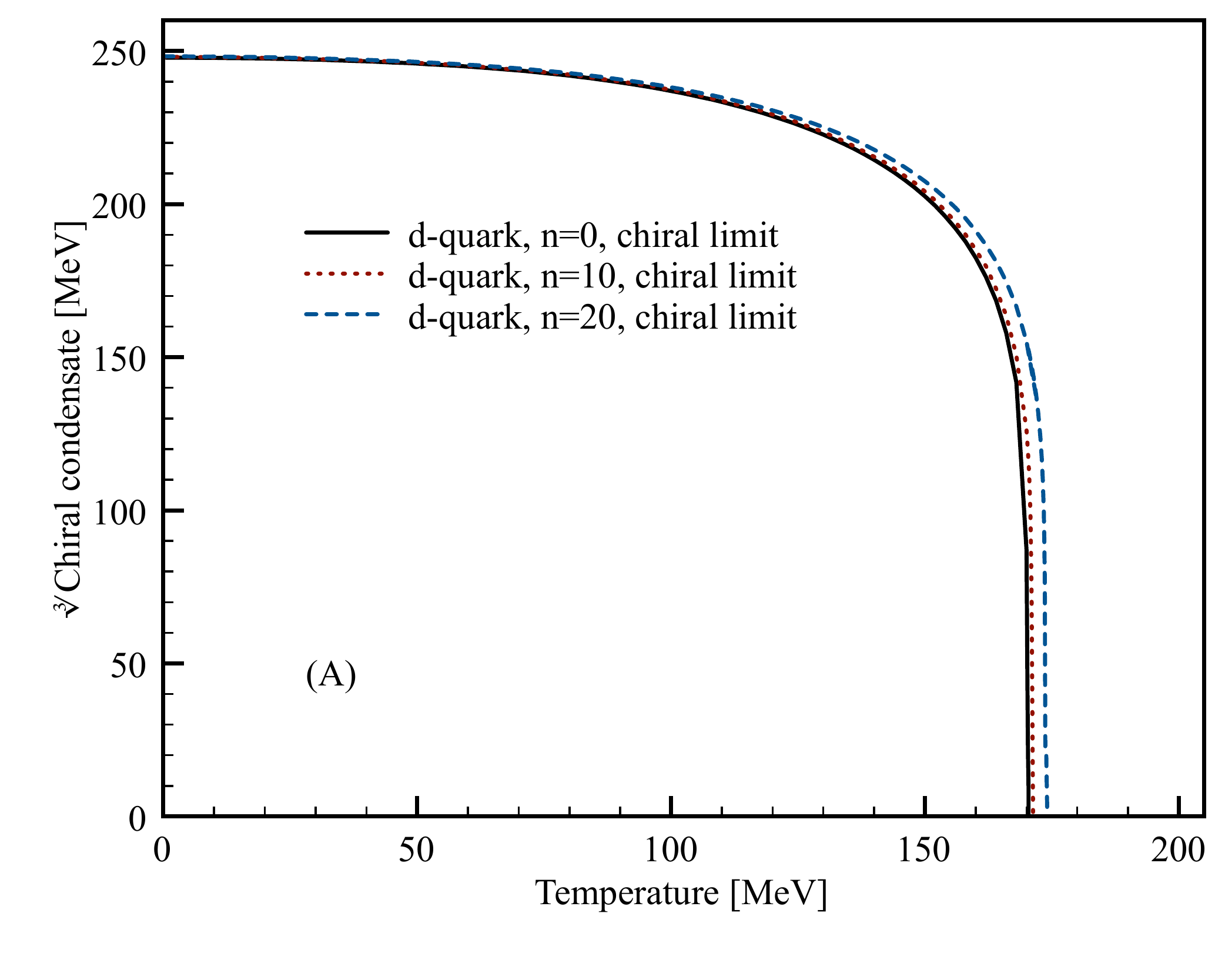}
\includegraphics[width=8.5cm]{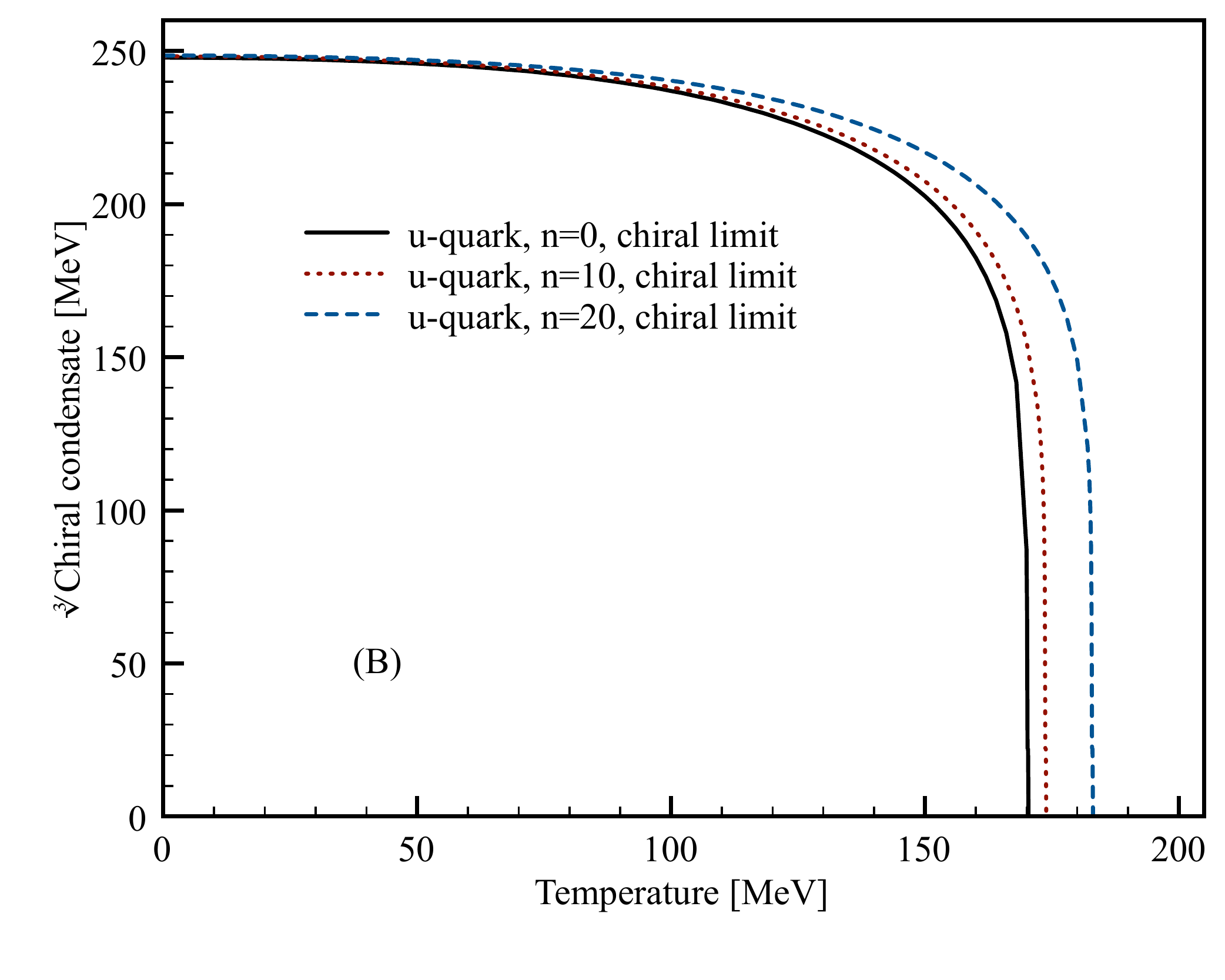}
\end{tabular}
\begin{tabular}{cc}
\includegraphics[width=8.5cm]{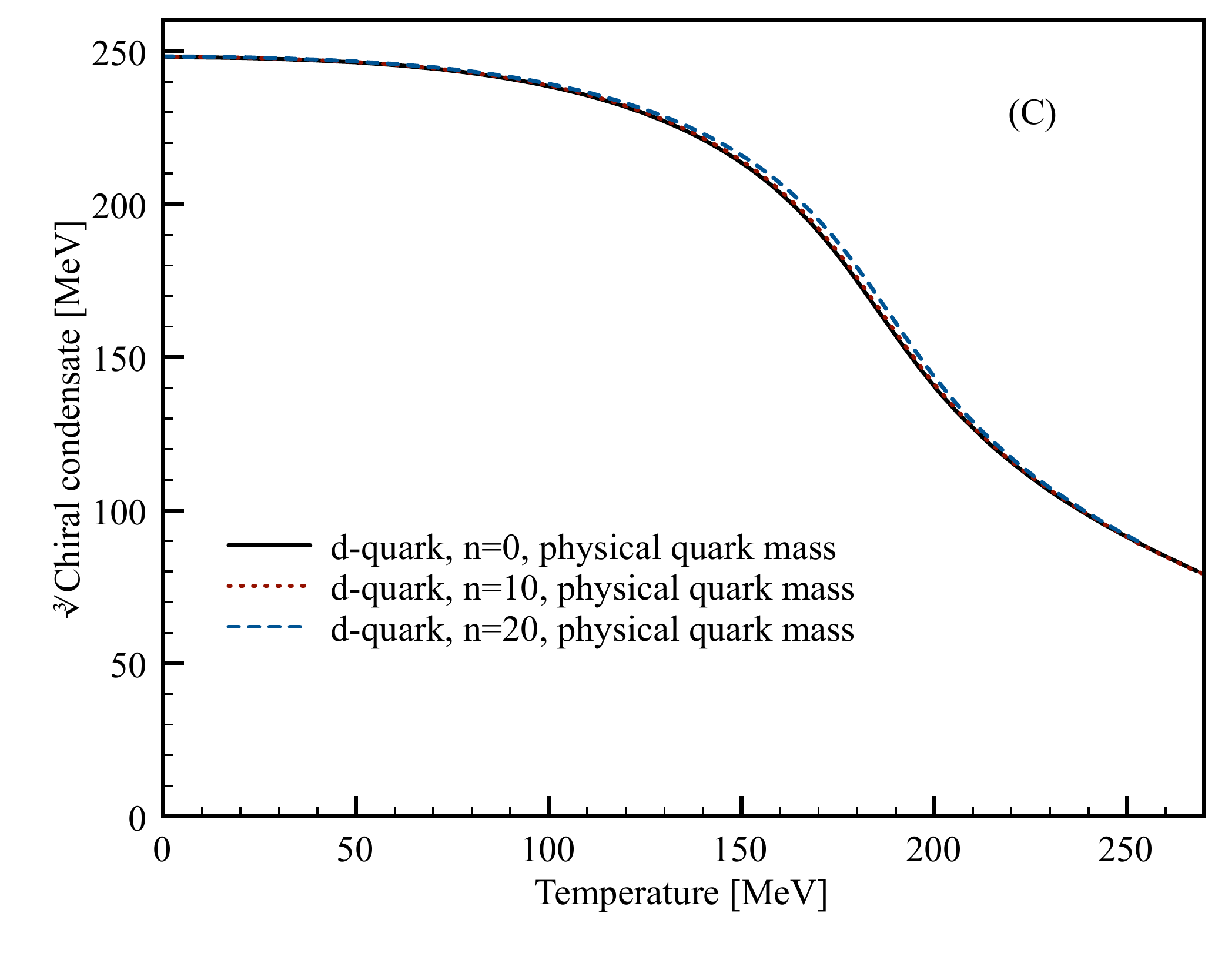}
\includegraphics[width=8.5cm]{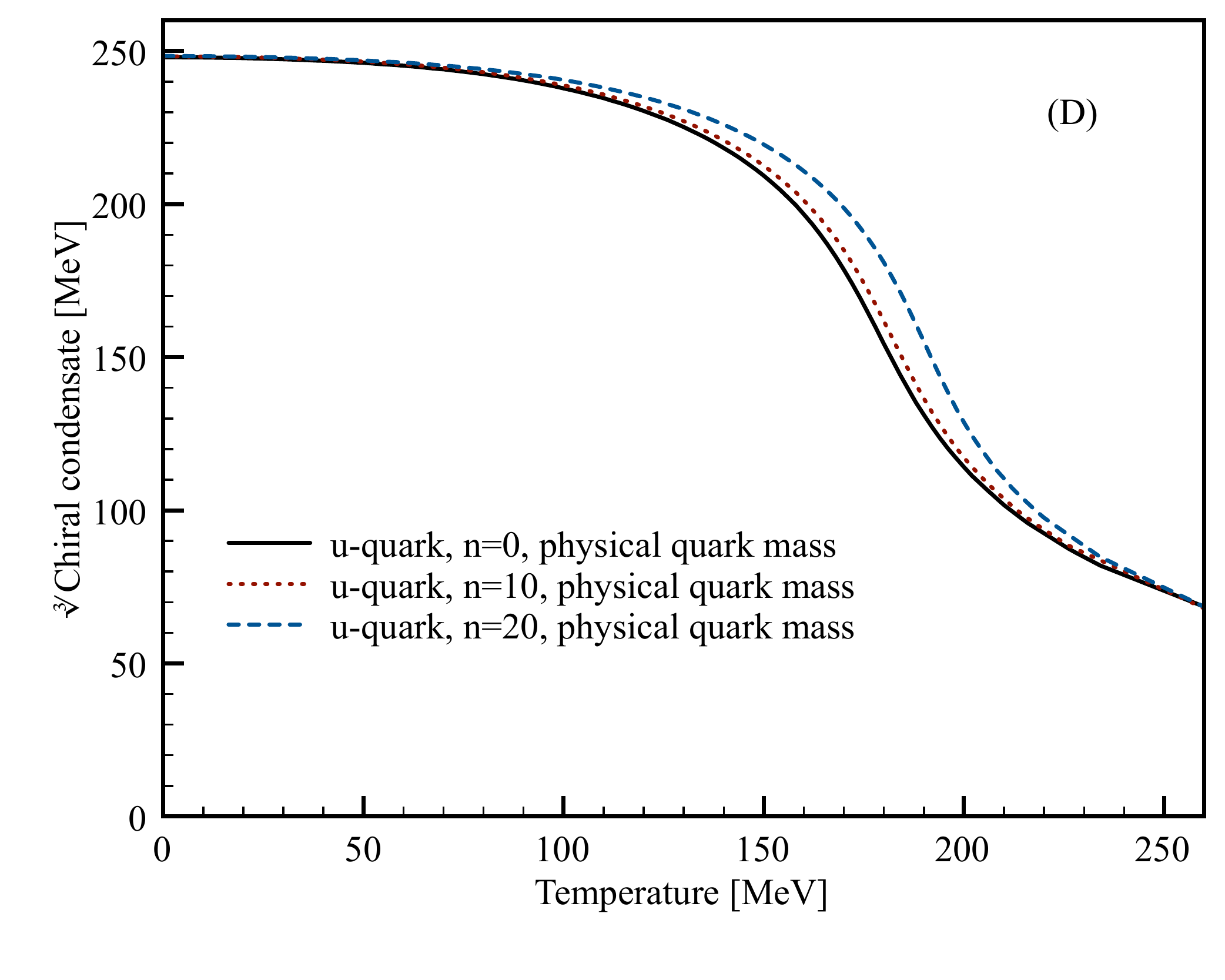}
\end{tabular}
\caption{Chiral condensate as a function of $T$. We draw the numerical results for the chiral limit and physical quark mass in the left and right columns. The results for the $d$ and $u$ quarks are given in the upper and lower rows, respectively. For more details, see the text. }
\label{FIG3}
\end{figure}

Now we are in a position to discuss the (partial) chiral restoration in the presence of the external magnetic field. It is indicated by the chiral order parameter, i.e. quark condensate. In Figure~\ref{FIG3}, we show the numerical results of the chiral condensate for the $d$ (left column) as well as $u$ (right column) quarks, separately.
We observe the second order chiral phase transition for the both flavors in the chiral limit case shown in the panel (A) and (B).
This result is expected from the universal class of the restoration pattern. Note that this correct restoration pattern in the present instanton framework is only achieved by the inclusion of the MLC as the large-$N_c$ corrections~\cite{Nam:2010mh}. Turning on the external magnetic field, one finds that the SB$\chi$S is enhanced, that is, the values for the quark condensate and $T_c$ both increase for the two flavor. Among them
we see that the $u$ quark condensate is more sensitive with respect to the magnetic field. It is
due to the larger quark electric charge of the $u$ quark.
The critical $T$ for the both flavors, $T^{u}_c$ and $T^{d}_c$ are listed in Table~\ref{TABLE1}. We observed that $\langle iu^\dagger u\rangle\approx\langle id^\dagger d\rangle\approx(247\,\mathrm{MeV})^3$ at $T=0$, which is just compatible to the empirical value of the isospin-symmetric quark condensate about $(250\,\mathrm{MeV})^3$.

Considering the physical quark mass case, the chiral phase transition for the two flavors are shown in the panel (C) and (D) of Figure~\ref{FIG3}. Following the universal class of the restoration pattern, the curves represent the crossover. The magnetic field effects are negligible for the $d$-quark condensate in the panel (C), due to the smaller quark electric charge, whereas $u$-quark condensate in the panel (D) shows visible changes in the vicinity of $T\approx180$ MeV with respect to the magnetic field. The $T_c$ can be obtained by computing the inflection point of the curves for the crossover phase transition~\cite{Rossner:2007ik}, resulting in that $T^{d}_c\approx200$ MeV for all the $n$ values and $T^{u}_c=(180\sim200)$ MeV for $n=(0\sim20)$. It is worthy of noting that the changes in the $T_c$ for the physical quark mass, due to the magnetic field, are relatively small in comparison to those for the chiral limit. This tendency is qualitatively consistent with the lattice QCD estimations~\cite{D'Elia:2010nq}. As for the physical quark mass case, the quark-condensate values for the both flavors at $T=0$ are about $(248\,\mathrm{MeV})^3$, which almost coincides with those for the chiral limit.

\begin{table}[h]
\begin{tabular}{c|c|c|c}
&$n=0$&$n=10$&$n=20$\\
\hline
$T^{u}_c$&$170.4$ MeV&$173.9$ MeV&$183.1$ MeV\\
$T^{d}_c$&$170.4$ MeV&$171.3$ MeV&$174.1$ MeV\\
\end{tabular}
\caption{Critical temperature for the $u$ and $d$ flavors, $T^{u}_c$ and $T^{d}_c$, for the chiral limit in the presence of the external magnetic field, $n=eB_0/m^2_\pi$.}
\label{TABLE1}
\end{table}

\subsection{Ratio of the two-flavor quark condensates: $\mathcal{R}$}
\begin{figure}[t]
\begin{tabular}{cc}
\includegraphics[width=8.5cm]{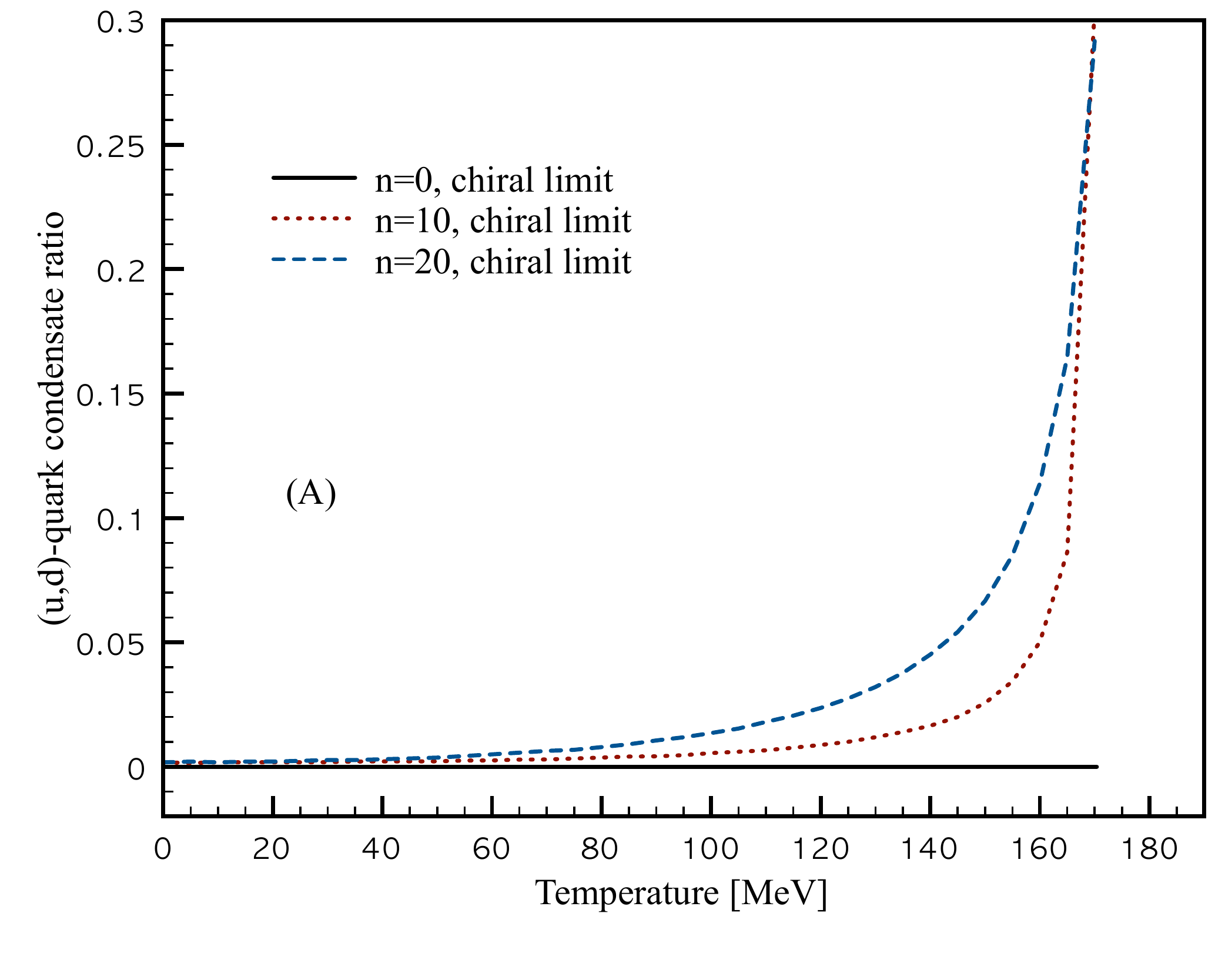}
\includegraphics[width=8.5cm]{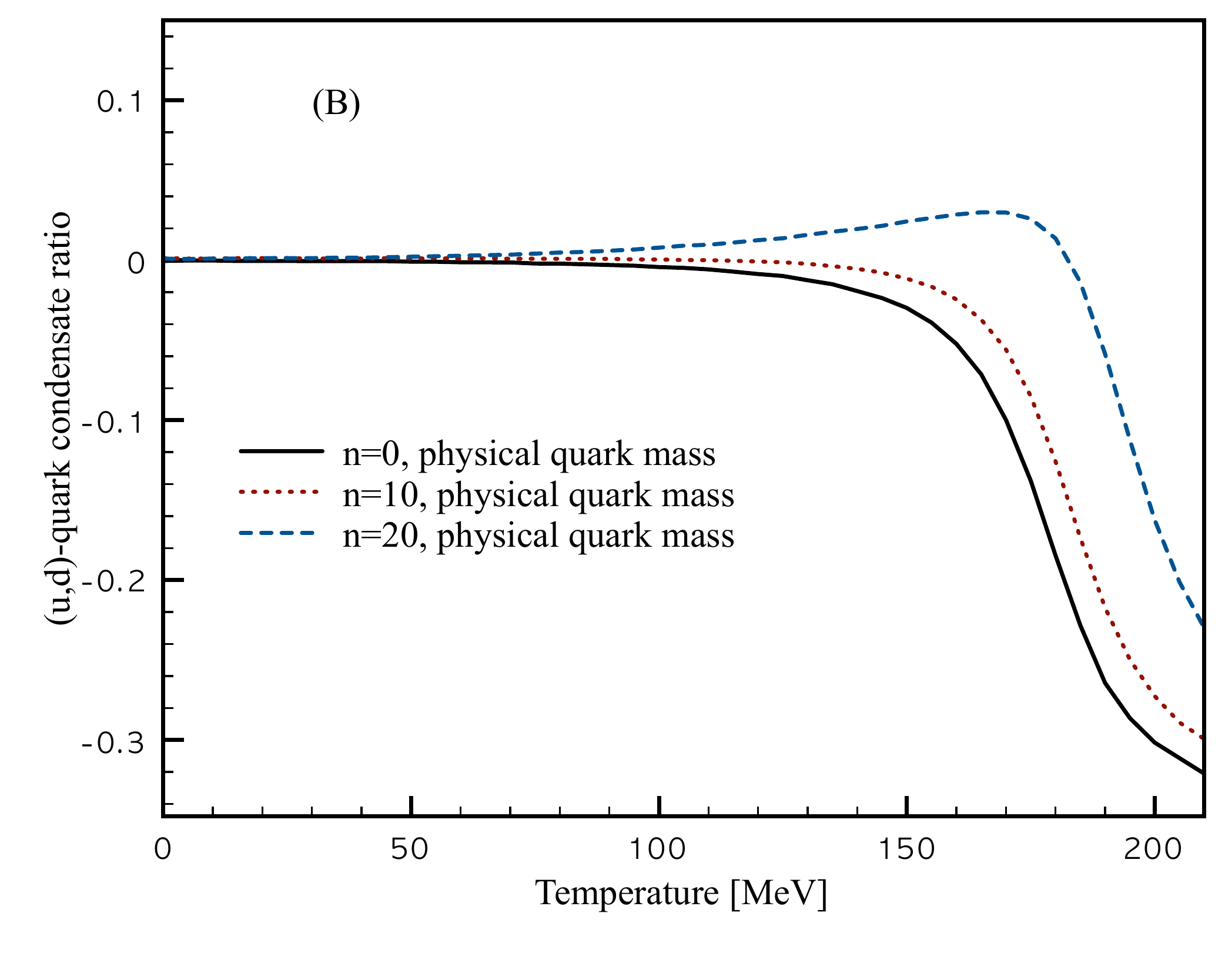}
\end{tabular}
\caption{$(u,d)$-quark condensate ratio $\mathcal{R}$ in Eq.~(\ref{eq:RATIO}) as function of $T$ for different $n\equiv eB_0/m^2_\pi$ values. We present the numerical results for the chiral limit and physical quark mass in the left and right panels, respectively.}
\label{FIG4}
\end{figure}

In the previous Subsections, we discussed the competition between the magnetic catalysis and diluting instanton effect at finite $T$, on top of the explicit isospin breaking. In the present Subsection, we want to take a more careful look on the isospin breaking of the quark condensates
by defining a quantity as in Eq.~(\ref{eq:RATIO}). We also note that the ratio $\mathcal{R}$ is deeply related to the low-energy constant of the $\chi$PT Lagrangian, $h_{3}$~\cite{Gasser:1983yg,Goeke:2010hm}. In Figure~\ref{FIG4}, we depict the $\mathcal{R}$ for the chiral limit in the panel (A) and physical quark mass case in the panel (B). In the chiral limit without the external magnetic field, the $u$- and $d$-quark condensates are the same so that $\mathcal{R}=0$ for any $T$ values. As the magnetic field increases from $n=0$, the $\mathcal{R}$ becomes a positive and stiffly increasing function. It is because that the $u$-quark condensate increases more rapidly than that for the $d$ quark and the magnetic catalysis effect is proportional to $e^2_f$ as in Eq.~(\ref{eq:CCT}). At the critical $T$, the values for the $\mathcal{R}$ diverges, signaling the second-order chiral phase transition.

The situation becomes quite different for the physical quark mass case in the panel (B). Without the magnetic field the $\mathcal{R}$  decreases then becomes negative beyond $T\approx100$ MeV.
As the magnetic field increases, the curves are shifted to higher $T$, and there appears a bump around $T=180$ MeV for $n=20$.
The apparent difference between the two cases is not hard to understand.
In the chiral limit, there is as a sort of intact degeneracy between $u$ and $d$-quark condensates. Such that
nonzero $\mathcal{R}$ values are only possible in the presence of the finite external magnetic field which breaks this degeneracy~\cite{Gusynin:1994va}. This is also true for the nonzero degenerated quark mass of the flavors, $m_u=m_d\ne0$.
On the contrary, if this degeneracy between the quark condensates is lifted up beyond $T\approx100$ MeV with decreasing nonperturbative effect (instanton), the explicit isospin breaking becomes more pronounced and results in the negative difference of the condensates at the zero magnetic field. It is due to the fact that the heavier quark causes the larger quark condensate in general. On the hand the $u$ quark is more sensitive to the magnetic catalysis because of its larger electric charge. In other words, the explicit isospin breaking effect pushes the $\mathcal{R}$ downward but
the magnetic catalysis pushes the $\mathcal{R}$ upward. This competition also causes a bump around $T=180$ MeV for the $\mathcal{R}$. Nevertheless $\mathcal{R}$ goes down when $T$ increases beyond $T=180$ MeV indicating the explicit isospin breaking effect due to the quark mass difference wins over magnetic catalysis there.

\subsection{Pion properties at finite $T$ under the magnetic field: $F_\pi$ and $m_\pi$}
Finally we want to make an analysis on the pion properties such as the pion weak-decay constant and pion mass, at finite $T$ in the presence of the external magnetic field, below the critical $T$. Since in the Nambu-Goldstone phase ($T$ is below $100$ MeV), the isospin symmetry is only slight broken as shown in Figure~\ref{FIG4}. Hence, in what follows, we focus on the properties with the isospin symmetry. For this purpose we employ the Gell-Mann-Oakes-Renner (GOR) relation defined as:
\begin{equation}
\label{eq:GOR}
m^2_\pi=\sum_{f=u,d}\frac{m_f}{F^2_\pi}\langle iq^\dagger_f q_f\rangle\to
\underbrace{\frac{2\bar{m}_f}{F^2_\pi}
\langle iq^\dagger q\rangle_{m_f=0}}_{\mathrm{isospin\,\,symmetric}}.
\end{equation}
Here, we defined $\bar{m}_f=(m_u+m_d)/2$. Since the quark condensates have been already computed as functions of $T$ as well as $B_0$ in the previous Sections, it is enough to calculate the pion-weak decay constant $F_\pi$ in the same framework. For simplicity, we ignore the MLC contribution to compute the $F_\pi$ and the nonlocal contribution for the time being~\cite{Nam:2008xx}. Then the analytical expression for the $F_\pi$ reads in the instanton framework for the vacuum as follows:
\begin{equation}
\label{eq:FPI1}
F^2_\pi\approx4\eta N_c\int\frac{d^4k}{(2\pi)^4}
\frac{M^2_k-k^2M_kN_k}{(k^2+M^2_k)^2},
\end{equation}
where we again have assumed the isospin symmetry. The $\eta$ denotes a  correction factor for the case without the nonlocal contribution. From Ref.~\cite{Nam:2008xx}, the value for the $\eta$ can be estimated as about $0.5$ to obtain the empirical value for the pion-weak decay constant, i.e. $F_\pi\approx93$ MeV. Although some dynamical information from the nonlocal contributions are missing by this simplification, it is still useful for a simple and qualitative analysis. If we induce the EM field externally, we can replace the constituent-quark mass squared approximately as $M^2_K\to M^2_k+2N^2_k\mathcal{B}^2_f$, according to Eq.~(\ref{eq:LO}). Moreover, taking into account that the term $M_kN_k$ can be obtained by differentiating $M^2_k/2$, we have $M_KN_K\approx M_kN_k+2N_k(\partial N_k/\partial k)\mathcal{B}^2_f$. Hence, we can write the expression for the $F_\pi$ as a function of $T$ and $B_0$, employing the fermionic Matsubara formula:
\begin{equation}
\label{eq:FPI2}
F^2_\pi\approx4\eta N_cT\sum_m\int\frac{d^3\bm{k}}{(2\pi)^3}
\frac{M^2_a-(\bm{k}^2+w^2_m)M_aN_a
+2N^2_a\bar{\mathcal{B}}^2_f}
{(w^2_m+\bm{k}^2+M^2_{\bm{k}})^2}=4\eta N_c\int\frac{d^3\bm{k}}{(2\pi)^3}
[\mathcal{K}_1+\mathcal{K}_2+\mathcal{K}_3\bar{\mathcal{B}}^2_f].
\end{equation}
Here, we have defined a flavor-averaged external magnetic field, i.e. the $\bar{\mathcal{B}}^2_f\equiv \frac{1}{2}\left(\mathcal{B}^2_u+\mathcal{B}^2_d \right)$, considering the isospin symmetric matter. Analytic expressions for the relevant functions $\mathcal{K}_{1\sim3}$ are given in the Appendix.

In Figure~\ref{FIG5}, we preset the numerical results of the pion weak-decay constant $F_\pi$ (A) and pion mass $m_\pi$ (B) as functions of $T$ and the strength of the magnetic field. In our numerical calculations, we have chosen $2\bar{m}_f\approx10$ MeV in Eq.~(\ref{eq:GOR}) as a trial, although we evaluated the analytical expression for the $F_\pi$ near the chiral limit. In the panel (A), the $F_\pi$ smoothly decreases with respect to $T$ indicating the partial chiral restoration. At $T=100$ MeV, the value of the $F_\pi$ is about $10\%$ reduced.
Increasing the strength of the magnetic field one finds that the value of $F_\pi$ is enhanced by a few percent.
At $T=0$, we have $F_\pi=(93.47,93.81,94.16)$ MeV for $n=(0,10,20)$, respectively. The effect from the magnetic catalysis appears more important in the higher $T$ region.

The numerical results for the $m_\pi$ are given in the panel (B). As a signal for the partial chiral restoration,
the pion mass increases with respect to $T$ but decreases with respect to $B_{0}$.
In other words,
the enhancement of the SB$\chi$S due to the magnetic catalysis is quite small.
Only about $0.5$ MeV decease in the pion mass is observed for $n=(0\to20)$. Again, the magnetic catalysis plays an more significant role in the higher $T$ region.

\begin{figure}[t]
\begin{tabular}{cc}
\includegraphics[width=8.5cm]{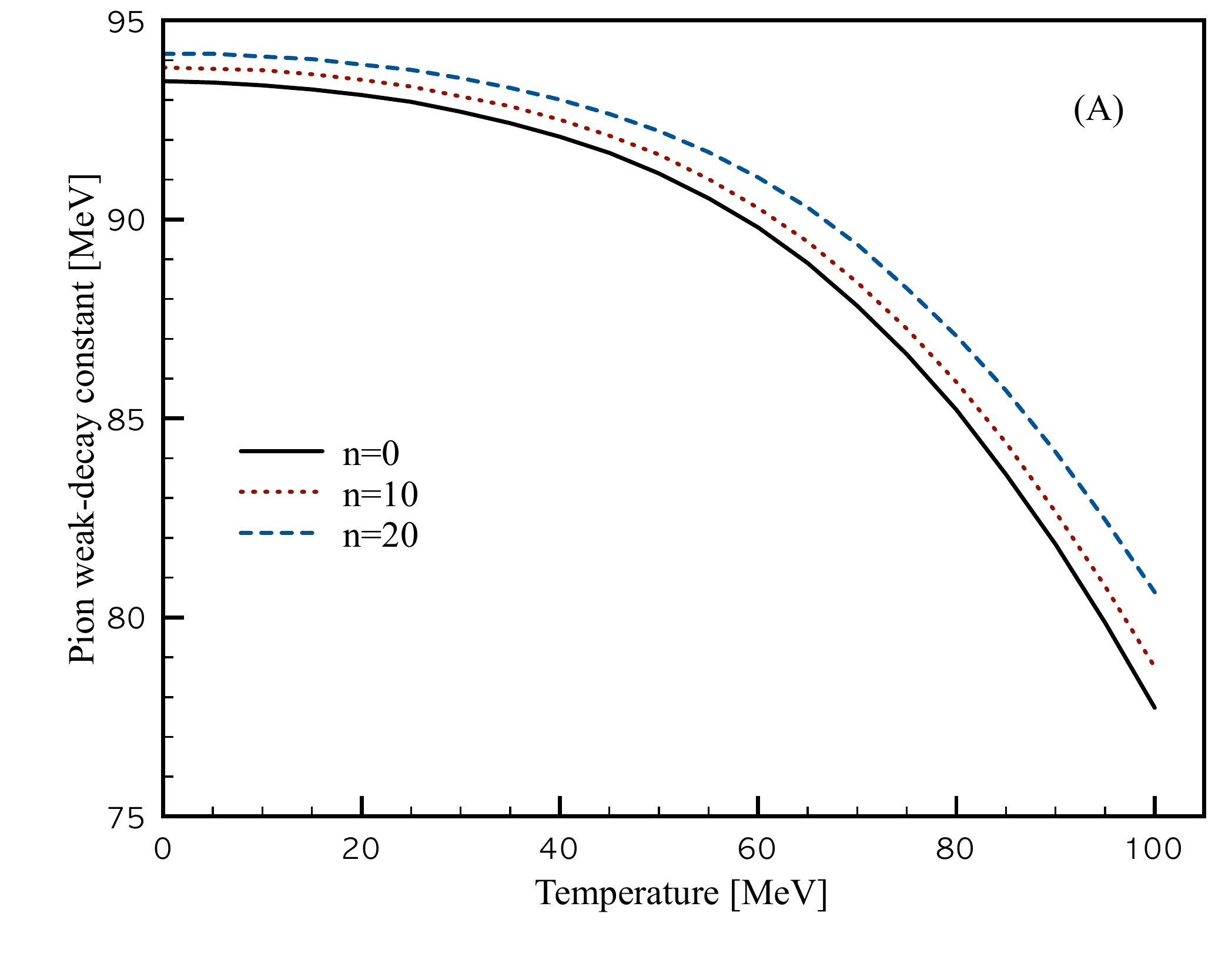}
\includegraphics[width=8.5cm]{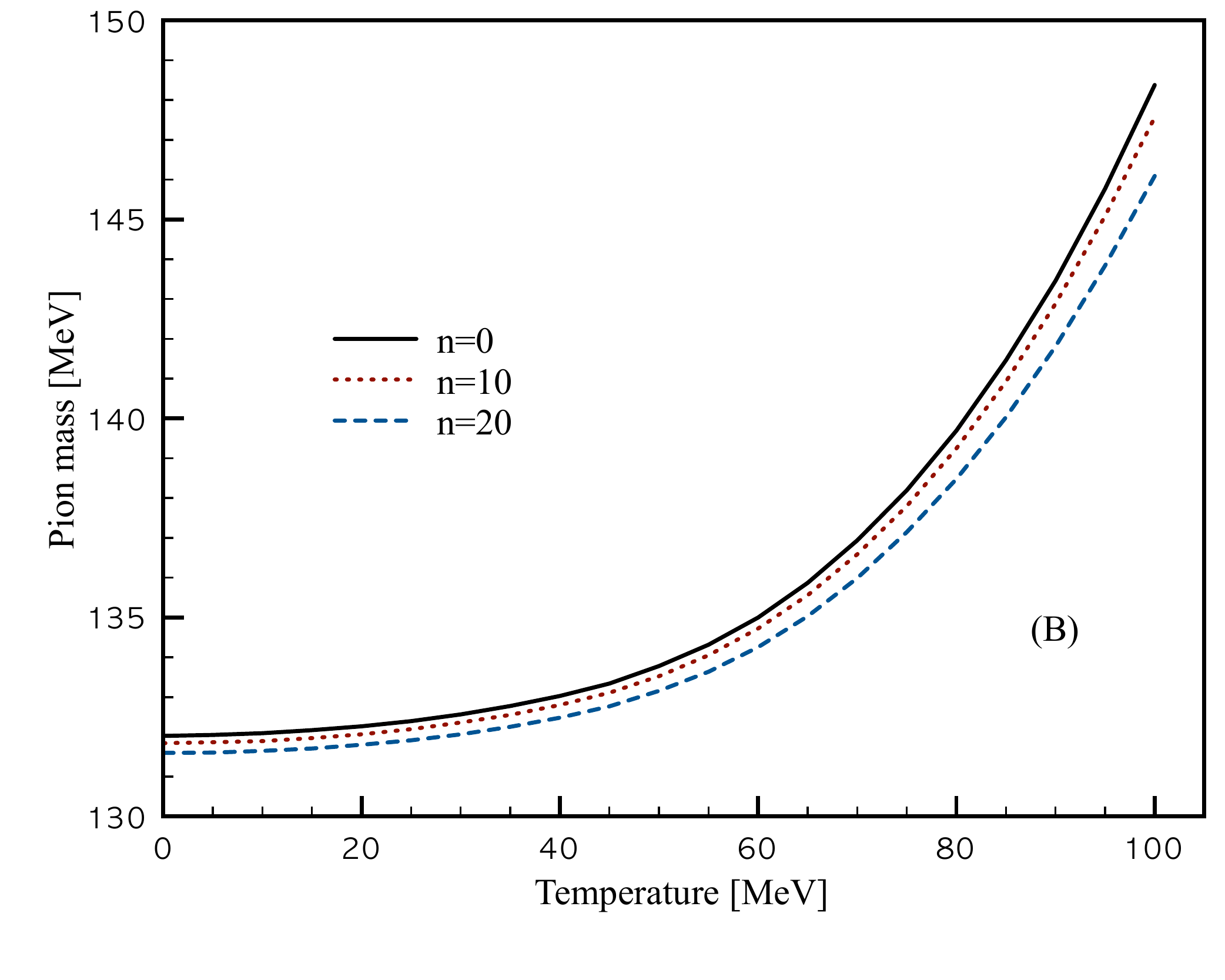}
\end{tabular}
\caption{Pion weak-decay constant $F_\pi$ (A) and pion mass $m_\pi$ (B) as a function of $T$, varying the strength of the magnetic field.}
\label{FIG5}
\end{figure}

\section{Summary and conclusion}
In the present work, we have investigated the (partial) chiral restoration at finite $T$ in QCD matter for the SU(2) light-flavor sector, under the strong and static external magnetic field. To this end, we employed the $T$-modified instanton-liquid model, together with the linear Schwinger method and fermionic Matsubara formula. We also took into the meson-loop corrections as the large-$N_c$ corrections to reproduce the correct phase transition pattern.
We then present the numerical results for the constituent-quark mass, chiral condensate, and isospin-symmetry breaking effect as functions of $T$, $B_0$, and flavor degrees of freedom. Below, we list important theoretical observations of the present work:
\begin{itemize}
\item Relevant instanton parameters $\bar{R}$ and $\bar{\rho}$ are modified as functions of $T$, resulting in the diluting instanton ensemble with respect to $T$, i.e. decreasing the SB$\chi$S effects. The external magnetic field enhances the SB$\chi$S in terms of the magnetic catalysis, which is proportional to $(e_fB_0)^2$. Hence, the $u$-quark constituent mass is more sensitive to the magnetic field and increases considerably more with respect to the field strength compared to the $d$ quark.
\item On top of the explicit isospin symmetry breaking, there appears a point at which the constituent quark masses for the $u$ and $d$ quarks coincide each other for the strong magnetic field $eB_0\approx10^{19}$ G. In the chiral limit, we observe the second-order chiral phase transition, as expected from the universal restoration pattern, effected much by the meson-loop corrections. Naturally, the crossover phase transition takes place for the physical quark masses.
\item The effects from the magnetic catalysis becomes more pronounced in the higher $T$ region because as $T$ increases the SB$\chi$S effects generated from the instanton is weakened so that the magnetic catalysis effects becomes relatively more important. The critical $T$, i.e. $T_c$ is shifted to higher $T$ due to the magnetic catalysis, whereas the change of the chiral condensate values is relatively small. $T_{c}$ becomes flavor-dependent because the magnetic catalysis effect depends on the electric charge of the quarks.
\item The isospin breaking between the quark condensates is explored by defining the ratio $\mathcal{R}$ as a function of $T$. As for $m_u=m_d$ case the ratio is zero at $B_0=0$ and monotonically increases with respect to $T$ for the finite magnetic field. For the physical quark mass case, the ratio $\mathcal{R}$ shows nontrivial structures with respect to $T$ and $B_{0}$ due to the complicated competition between the magnetic catalysis and the explicitly isospin breaking effect which becomes more important at higher T because of the decreasing nonperturbative effects.
\item According to our simple and qualitative analysis using the GOR relation, we observe correct partial chiral-restoration and magnetic-catalysis behaviors for the pion-weak decay constant $F_\pi$ and pion mass $m_\pi$. They decreases and increases about $10\%$ at $T\approx100$ MeV in comparison to those at $T=0$, respectively. However, the changes due to the magnetic field are relatively small, just a few percent.
\end{itemize}

Now we obtain an effective chiral action at finite $T$ as well as the magnetic field for the physical quark mass. If the strong magnetic field is created in the peripheral heavy-ion collisions as reported, it is worthy of studying the hadronization processes in the presence of the magnetic field, i.e. the dilepton production via the vector-meson dominance under the magnetic field for instance. Moreover, the QCD phase diagram on the $\mu$-$T$ plane and critical values, such as the critical end point (CEP) and tricritical point (TCP), are also able to be explored in our model in principle. Inclusion of finite $\mu$ to the effective action in the instanton framework is under progress and related works will appear elsewhere.
\section*{Acknowledgment}
The authors are grateful to B.~G.~Yu for fruitful discussions. The work of S.i.N. was supported by the grant NRF-2010-0013279 from National Research Foundation (NRF) of Korea. The work of C.W.K. was supported by the grant NSC 99-2112-M-033-004-MY3 from National Science Council (NSC) of Taiwan. C.W.K has also acknowledged the support of NCTS (North) in Taiwan.
\section*{Appendix}
The relevant functions in Eq.~(\ref{eq:NOVT}) and (\ref{eq:CCT}) are given as follows:
\begin{eqnarray}
\label{eq:FFFF}
\mathcal{F}_1&=&M_a\bar{M}_a\mathcal{H}_2,
\,\,\,\,
\mathcal{F}_2=2N^2_a\mathcal{H}_2,
\,\,\,\,
\mathcal{F}_3=M_aM_b\mathcal{H}_4
+M_aM_b\left[\bm{k}\cdot(\bm{k}+\bm{q})\right]\mathcal{H}_3,
\cr
\mathcal{F}_4&=&M_aM_b
\left[\bar{M}_{a}\bar{M}_{b}+M_{a}M_{b}+\frac{m_f}{2}(M_{a}+M_{b})\right]
\mathcal{H}_3
\cr
\mathcal{G}_1&=&\bar{M}_a\mathcal{H}_2,\,\,\,\,\mathcal{G}_2=-m_f\mathcal{H}_1,
\,\,\,\,
\mathcal{G}_3=M_aM_b(\bar{M}_a\bar{M}_b)\mathcal{H}_3,
\cr
\mathcal{K}_1&=&(M^2_a-\bm{k}^2M_aN_a)\mathcal{H}_5,\,\,\,\,
\mathcal{K}_2=-M_aN_a\mathcal{H}_6,\,\,\,\,
\mathcal{K}_3=2N^2_a\mathcal{H}_5. \nonumber
\end{eqnarray}
$\mathcal{H}_{1\sim 6}$ are explicitly given as follows,

\begin{eqnarray}
\label{eq:RF2}
\mathcal{H}_{1}&=&T\sum_{m}\frac{1}{w^{2}_{m}+E^{2}_{0}}
=\frac{1}{2E_{0}}\mathrm{tanh}\left(\frac{E_{0}}{2T} \right),\,\,\,\,
\mathcal{H}_{2}=T\sum_{m}\frac{1}{w^{2}_{m}+E^{2}_{a}}
=\frac{1}{2E_{a}}\mathrm{tanh}\left(\frac{E_{a}}{2T} \right),
\cr
\mathcal{H}_{3}&=&T\sum_{m}
\frac{1}{(w^{2}_{m}+E^{2}_{a})(w^{2}_{m}+E^{2}_{b})}
=\frac{1}{2E_{a}E_{b}(E^{2}_{a}-E^{2}_{b})}
\left[E_{a}\mathrm{tanh}\left(\frac{E_{b}}{2T} \right)
-E_{b}\mathrm{tanh}\left(\frac{E_{a}}{2T} \right) \right],
\cr
\mathcal{H}_{4}&=&T\sum_{m}
\frac{w^2_n}{(w^{2}_{m}+E^{2}_{a})(w^{2}_{m}+E^{2}_{b})}
=\frac{1}{2(E^{2}_{a}-E^{2}_{b})}
\left[E_{a}\mathrm{tanh}\left(\frac{E_{a}}{2T} \right)
-E_{b}\mathrm{tanh}\left(\frac{E_{b}}{2T} \right) \right],
\cr
\mathcal{H}_5&=&T\sum_{m}\frac{1}{(w^{2}_{m}+E^{2}_{a})^2}
=\frac{1}{8TE^3_a}\mathrm{sech}^2\left(\frac{E_a}{2T} \right)\left[T\,\mathrm{sinh}\left(\frac{E_a}{T} \right)-E_a \right],
\cr
\mathcal{H}_6&=&T\sum_{m}\frac{w^2_n}{(w^{2}_{m}+E^{2}_{a})^2}=
\frac{1}{8TE_a}\mathrm{sech}^2\left(\frac{E_a}{2T} \right)\left[E_a+T\,\mathrm{sinh}\left(\frac{E_a}{T} \right)\right].
\nonumber
\end{eqnarray}

\end{document}